\documentclass[reprint,showpacs,amsmath,amssymb,aps]{revtex4-1}
\usepackage{graphicx}
\usepackage{amsmath}
\usepackage{rotfloat}
\usepackage[colorlinks,linkcolor=red,anchorcolor=blue,citecolor=green]{hyperref}
\usepackage{cleveref}
\usepackage{natbib}

\def\BibTeX{{\rm B\kern-.05em{\sc i\kern-.025em b}\kern-.08em
		T\kern-.1667em\lower.7ex\hbox{E}\kern-.125emX}}

\begin{document}
\title{Thermalization Effect in semiconductor Si, and metallic silicide NiSi$_{2}$, CoSi$_{2}$ by using Non-Adiabatic Molecular Dynamics Approach}
\makeatletter
\let\@fnsymbol\@arabic
\makeatother
\author{Kun Luo$^{1,2,\dag}$
Weizhuo Gan$^{1,2,\dag}$
Zhaozhao Hou$^{3}$
Guohui Zhan$^{1,2}$
Lijun Xu$^{1,2}$
Jiangtao Liu $^{4}$\footnote{jtliu@semi.ac.cn}
Ye Lu$^{5}$ \footnote{lu_ye@fudan.edu.cn}
Zhenhua Wu$^{1,2}$ \email{wuzhenhua@ime.ac.cn}}

\affiliation{$^1$ Key Laboratory of Microelectronics Devices and Integrated Technology, Institute of Microelectronics, Chinese Academy of Sciences, Beijing 100029, China}
\affiliation{$^2$ School of Integrated Circuits, University of Chinese Academy of Sciences, Beijing 100049, China}
\affiliation{$^3$ HiSilicon Technologies, Shenzhen, China}
\affiliation{$^4$ College of Mechanical and electrical engineering, Guizhou Minzu University, Guiyang 550025, China}
\affiliation{$^5$ The School of Information Science and Technology, Fudan University, Shanghai 200433, China}
\affiliation{$\dag$ The authors contribute equally}
\email{wuzhenhua@ime.ac.cn; lu_ye@fudan.edu.cn; jtliu@semi.ac.cn.}


\begin{abstract}
Recently, cold source transistor (CSFET) with steep-slope subthreshold swing (SS) $< 60$ $mV/decade$ has been proposed to overcome Boltzmann tyranny in its ballistic regime.
However the scattering, especially by inelastic scattering may lead serious SS degradation through cold carrier thermalization.
In this study, the electronic excitation/relaxation dynamic process is investigated theoretically by virtue of the state-of-the-art nonadiabatic molecular dynamics (NAMD) method, i.e., the mixed quantum-classical NAMD.
The mixed quantum-classical NAMD considers both carrier decoherence and detailed balance to calculate the cold carrier thermalization and transfer processes in semiconductor Si, and metallic silicide (NiSi$_{2}$ and CoSi$_{2}$).
The dependence of the thermalization factor, relaxation time, scattering time and scattering rate on energy level are obtained.
The thermalization of carrier gradually increases from low energy to high energy.
Partially thermalization from the ground state to reach the thermionic current window is realized with sub-100 $fs$ time scale.
Fully thermalization to entail energy region depends on the barrier height sensitively, i.e., the scattering rate decreases exponentially as the energy of the out-scattering state increase.
The scattering rate of NiSi$_{2}$  and CoSi$_{2}$  is 2 orders of magnitude higher than that of Si, arising from their higher density of states than that in Silicon
This study can shed light on the material design for low power tunneling FET as well as the emerging CSFET.
\end{abstract}


\maketitle

\section{Introduction}

The development of integrated circuits following the Moore's law promotes the traditional information industry, as well as the artificial intelligence, Internet to thins, big data and other emerging fields.
As the devices are miniaturized and integrated, their physical dimensions are getting closer and closer to the physical limits.
In addition, there also exist be a power consumption wall due to scaling limit of supply voltage ($V_{DD}$) with sufficient on/off ratio, becomes one critical issue in transistor technology~\cite{datta2022}.
To overcome this limit, new transistor switching mechanisms are investigated to achieve a steep subthreshold swing (SS), such as tunnel FETs (TFETs)\cite{seabaugh2010,ionescu2011,lu2014,guo2020PRL}, negative-capacitance FETs (NCFETs)\cite{Salahuddin2008, Ajay2019,guo2021}.
Recently a novel steep slop CSFETs~\cite{cheung2010,fliu2018}, including Dirac source FET as a special case of cold source FET~\cite{fliu2018-science,liu2020IEDM,pzhou2021nanolett,liu2021ACSNano}, has been proposed.
Various designs of density of states engineering scheme can filter cold carriers in the source.
Sub 60 mV/decade SS is predicted, in different material systems such as graphene/MoS$_{2}$,\cite{jzchen2020,wang2021}, Graphene/Cd3C2/T-VTe2/H-VTe2/H-TaTe2-InSe~\cite{lyu2020}, metal Heusler alloy-WS$_{2}$/MoSe$_{2}$~\cite{liu2020PRApp}, KTlCl$_{4}$/Au$_{2}$S~\cite{logoteta2020}
, artificial type-III heterojucntion of silicon~\cite{gan2020ted,gan2021ted,qi2022}, et.al., without considering any scattering.

In path-finding stage, most kinds of steep-slope CSFETs have been predicted by non-equilibrium Green's function method with density function theory (NEGF/DFT) in ballistic transport regime, except for the Dirac source FETs that have been demonstrated experimentally~\cite{fliu2018-science,liu2020IEDM,pzhou2021nanolett,liu2021ACSNano}.
The carrier transport mechanisms and the predicted performance of versatile cold source FET have not been addressed thoroughly yet.
In advanced ultra scaled transistor, even if the transport length is less than 20 nm, the performance will be degraded by various scatters \cite{svizhenko2003role,stanojevic2015,yao2018}.
Normally, the scattering processes can be divided elastic scattering and inelastic scattering according to whether the energy changes.
Elastic scattering such as coulomb scattering changes the carrier momentum distribution but does not change the energy distribution, so it has an effect on carrier mobility but has no direct effect on carrier energy relaxation.
On the other hand, inelastic scattering, such as electron-phonon scattering and electron-electron scattering, will lead to the transition of cold carriers to high energy level occupied states, i.e., rethermalization.

It is getting increasingly important to understand the thermal, electrical and thermoelectric transport properties of materials accounting for the scattering.
There have been several reports on the first-principles electron-phonon interactions calculation and carrier relaxation time estimation with high accuracy \cite{jzhou2016,jap2018}.
For large systems it is computational expensive to follow the above first-principle method.
Alternatively, NAMD approach~\cite{crespo2018} is widely used to study carrier dynamic processes.
To further release the simulation burden, there are several NAMD algorithms carried out in a mixed quantum-classic (MQC) fashion, e.g., Mean-field Ehrenfest dynamics (MFE) \cite{ren2013,parandekar2005mixed} and fewest switches surface hopping (FSSH) \cite{tully1990,drukker1999}.
Although both of them are simpler and more widely used algorithms, there are several issues because the difference between quantum and classical mechanics.
MFE method lacks of the detailed balance and does not include the quantum-mechanical wave function of the phonon or any electronic coupling.
This means that MFE is hard to study equilibrium properties and energy relaxation processes.
For FSSH method, the original surface-hopping algorithm does not have the proper decoherence.
Recently, a new NAMD algorithms with introducing a state density matrix formalism is proposed by Kang, et.al.
Under the modified conventional density matrix approach, MQC-NAMD method can incorporate the detailed balance and decoherence effect~\cite{kang2019nonadiabatic}.
Note that in such large systems, the nuclear movement is not conspicuously changed by single electron wave function, it is crucial to study on the carrier dynamics rather than the nuclear movement~\cite{akimov2013}, which is enough for the problem of hot-carrier cooling or cold-carrier excitation.
With this state-of-the-art NAMD approach, it is possible to calculate carrier dynamics for large systems with hundreds of atoms at the first-principle density functional theory level with high accuracy and efficiency.

\begin{figure}[!t]
	\centering{\includegraphics[width=8.5cm]{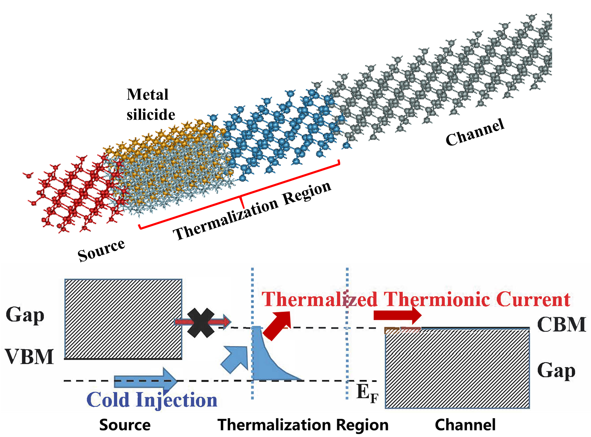}}
	\caption{(a) Schematic of atomic structure of Silicon CSFET. (b) Current transport diagram of cold source transistor in off state. Taking N-type transistor as an example, a P-type semiconductor band gap truncation is considered as an electron cold source, and the current on the channel barrier is injected through the thermal zone.}
	\label{ThermalTransport}
\end{figure}

Most researches on carrier dynamics with NAMD method focus on the hot carrier relaxation~\cite{kang2019nonadiabatic}, transfer~\cite{wang2019jpcl}, and cooling,\cite{prl1981,nc2017,banerjee2020}
The carrier dynamics of the reverse process, excitation from ground state to high energy states, has been seldom reported.
It can be a crucial factor affect the performance of tunneling FET and CSFET.
In the present work, we focus on the thermalization process, so as to get more physical insight on the carrier transport in Silicon cold source FET~\cite{gan2020ted}.
In Silicon CSFET, an artificial type-III heterojucntion of p-Si/Metal/n-Si in source is designed to achieve density of states engineering.
The bandgap in Source I (see Fig.\ref{ThermalTransport}) cuts off high-energy carriers; metallic silicide layer, usually being NiSi$_{2}$  and CoSi$_{2}$, is introduced to reduce the tunneling barrier and thus boost the driven current.
To investigate the carrier thermalization process in Si CSFET, the state-of-the-art NAMD method is used to study inelastic scatterings, such as electron-phonon scattering~\cite{giannini2018JPCL}, in semiconductor Si and two type of metallic silicides (NiSi$_{2}$  and CoSi$_{2}$).
The dependence of the thermalization factor, relaxation time, scattering time and scattering rate on energy level are obtained.
The relationship established here enables us to not only predict the thermalization time for cold carriers in semiconductors, metallic silicides, but also provide insights into determinant factors of thermalization.

\section{Methodology}
In the adiabatic approximation, i.e., the Born-Oppenheimer (BO) approximation, the nucleus is regarded as fixed, and the electron and nuclear motions are separated, since the mass of the nucleus is three orders of magnitude greater than that of the electron, the electron moves much faster than the atom.
An electron is considered to exhibit certain energy level properties within the potential field of the surrounding atoms at fixed positions.
In DFT calculations, the lattice temperature is generally set to 0 K, while only the electronic temperature is set, which is convenient for solving the Kohn-Sham equation in a fixed atomic position system.
However, many important physical phenomena involve the dynamics of excited states of carriers, such as the cooling of hot carriers including the photo-generated hot carrier cooling \cite{banerjee2020effects,bernardi2015ab}, hot carrier injection in transistors \cite{takeda1983empirical}, and electron-hole recombination \cite{chu2020low}.
Therefore, at a specific temperature, the real-time change of atomic position by lattice vibration and the dynamic process of energy and momentum interaction between electrons and phonons with electron-phonon coupling need to be considered in appropriate methods.
Lattice vibrations can be described by a combination of a series of harmonic oscillators.
However at room temperature, the phonons of many materials have an non-harmonic components, which can be considered by Molecular dynamics (MD) simulations \cite{markussen2017electron}.
For the cold carrier thermalization process, the energy of the cold carrier may be changed again by electron-phonon scattering or electron-electron scattering in the thermalization region.
And it is necessary to consider that its distribution extends from a narrow low energy range to the high energy levels to finally reach equilibrium at a finite time.
Through the non-adiabatic calculation process, the scattering of electron-phonon coupling on the carrier distribution can be introduced, so as to to calculate the evolution process with time dependent carrier distribution.
Note that electron-electron scattering is not included in this study. It is good for cases where the concentration of excited carrier is small as decreasing exponential with energy, so carrier-carrier interaction can be ignored.
Non-adiabatic molecular dynamics (NAMD) is combined with semi-classical and semi-quantum calculation method \cite{barbatti2007fly,crespo2018recent}, in which the nuclear motion and carrier dynamics are calculated separately.
The electrons are calculated by the time-dependent Schr$\ddot{o}$dinger equation, while the motion of the nucleus is calculated by Newton's second law.
First, the nuclear motion is calculated by adiabatic molecular dynamics (BOMD, Born-Oppenheimer molecular dynamics). One can use the first principle combined with molecular dynamics calculation to obtain the nuclear motion orbit as well as the wave function and energy of the adiabatic state.
The adiabatic state in BOMD has its electronic structure determined by the position of the atomic nucleus, and the effect of the electronic motion on the atomic nucleus is neglected, which is called the classical path approximation \cite{akimov2013pyxaid}.
For the system which mainly focuses on the carrier dynamics process and whose carrier excited state has little influence on the nuclear motion, the simulation of large time and space scale can be calculated.
So that DFT calculation can be carried out on the structure of hundreds of atom, and then the carrier dynamics process can be simulated.
The time evolution of the wave function and the overlap between the wave functions of the adiabatic states can be obtained from the BOMD, which reflects the influence of the lattice vibration on the electronic structure.
On this basis, the time evolution of the carrier excited States can be obtained from the NAMD calculation of the non-adiabatic process.
At each time, the adiabatic state $\Phi_i(t)$ and eigenenergy $\varepsilon_i(t)$ are determined by the Hamiltonian of single electron $H(t)$, and the non-adiabatic state $\Psi(t)$ can be expanded by each adiabatic states which are satisfied the time-dependent Schr$\ddot{o}$dinger equation,
\begin{equation}
H(t)\Phi_i(t)=\varepsilon_i(t)\Phi_i(t)
\end{equation}
\begin{equation}
\Psi_l(t)=\sum\limits_{i}C_i^l(t)\Phi_i(t)
\end{equation}
The charge density of the collection of electrons in the system can be expressed by the density matrix as,
\begin{equation}
	\rho(r)=\sum_{i,j}D_{i,j}\Phi_i(r)\Phi_j(r)^*
\end{equation}
\begin{equation}
	D_ij(t)=\sum_l\omega_l(t)C_i^l(t)C_j^l(t)^*
\end{equation}	
where $\omega_l(t)$ is the statistical weight of $\Psi_l(t)$. Decoherence effect and detailed balance condition can be introduced by matrix of P, and the equation of density matrix variation with time can be written as \cite{kang2019nonadiabatic},
\begin{equation}
	\begin{split}
		\frac{\partial}{\partial t}D_{ij}(t)
		& =-i\sum_k[V_ik(t)D_kj(t)-D_ik(t)V_kj(t)]\\
		& -(1-\delta_{ij})\frac{D_{ij}(t)}{\tau_{ij}(t)}\\
		& =-i[V,D]_{ij}-(1-\delta_{ij})\frac{D_{ij}(t)}{\tau_{ij}(t)}\\
	\end{split}
\label{Dij}
\end{equation}
\begin{equation}
	V_{ij}(t)=\delta_{ij}\epsilon_i(t)-i<\Phi_i(t)|\frac{\partial \Phi_j(t)}{\partial t}
\end{equation}
where $\delta_{ij} $ is delta function and $\tau_ij(t)$ is the decoherence time.
The second term in Eq.~\ref{Dij} introduces a decay term for the off-diagonal elements of the density matrix to account for decoherence effects. While the first term $V_ij$ includes the change of the electron wave function with time under the electron-phonon coupling. In order to import a particular balance condition to ensure that the system tends to the Boltzmann distribution after a long time interaction, the density matrix $D_{ij}$ is split to $D_{ij}=P_{ij}+P_{ij}^*, (P_{ij} \neq P_{ij}^*)$, where $P_{ij}$ describes the transformation of an electron from $j$ state to $i$ state. The diagonal elements of the density matrix $D_ii=2P_ii$ describes the change in charge density,
\begin{equation}
	\begin{split}
		\frac{\partial}{\partial t}P_{ii}
		& =-Re(i[V,P]_{ij})\\
		& +\sum_jRe(iP_{ij}V_{ij})f_{ij}(exp(\frac{-|\Delta\epsilon_{ij}|}{k_{B}T})-1)\\
		& -\sum_jRe(iP_{ij}V_{ij})(1-f_{ij})(exp(\frac{-|\Delta\epsilon_{ij}|}{k_{B}T})-1)\\
	\end{split}
\end{equation}
where $\Delta\epsilon_{ij}=\epsilon_i-\epsilon_j$.
For electron scattering, when $\Delta\epsilon_{ij} > 0$, $f_{ij}$ is equal to 1, when $\Delta\epsilon_{ij} < 0$, $f_{ij}$ is equal to 0.
If the energy of the final state is higher than that of the initial state, it is limited by the Boltzmann factor $exp(\frac{-|\Delta\epsilon_{ij}|}{k_BT})$.
Otherwise, If the energy of the final state is lowerr than that of the initial state, there is no limitation.
Both the decoherence effect and the detailed balance condition can be introduced into the NAMD calculation through the $P$-matrix approach.
The decoherence time \cite{wong2002solvent} can be given by direct calculation or as a input parameter.\\
In this work, we employ the NAMD module in PWmat package\cite{jia2013analysis} to study the cold carrier thermalization process in semiconductor Si and metallic silicide (NiSi$_{2}$  and CoSi$_{2}$).
The calculation is divided into two steps.
First, the Hamiltonian of the ground state electron is calculated by BOMD to obtain the wave function and the eigenenergy.
Meanwhile, the overlapping of the adiabatic states at a series of times is output under a selected nonadiabatic energy window, which can present as $S_{ij}=<\Phi_i(t)|\Phi_j(t+dt)>$.
And the initial vibration of each atom is randomly generated in BOMD according to the temperature via $k_{B}T$.
The second step is to calculate non-adiabatic dynamics process of carries.
The evolution process of the carrier in the non-adiabatic window with time is investigated by selecting the initial time, the initial occupation distribution and the range of the non-adiabatic energy window.

\section{Results and Discussions}
\subsection{The carrier thermalization process in bulk silicon}
\subsubsection{Crystalline structure of silicon}
\label{subsubsec:Si}
\begin{figure}[!t]
	\centering{\includegraphics[width=8.5cm]{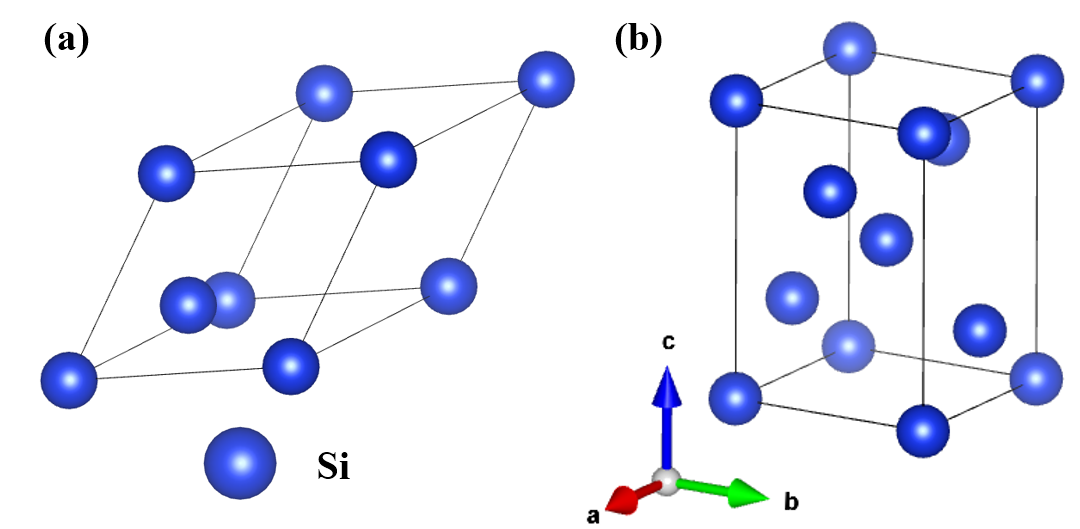}}
	\caption{Schemetic of bulk Silicon. (a) primitive cell structure with lattice constant of 3.840 \AA, (b) unit cell taken along [1 1 0], [1 -1 0], [0 0 1] directions. The volume of the primitive cell and the unit cell are 40.04 \AA and 80.08 \AA, respectively. And the c axis direction is [0 0 1] }
	\label{Siprim}
\end{figure}
First, the electronic structure and density of states of the bulk Si are calculated.
The primitive cell of diamond-structure bulk Si is shown in Fig.\ref{Siprim}(a), and the lattice constant is 3.840 $\AA$.
The norm-conserving pseudopotential (NCPP) and Heyd-Scuseria-Ernzerhof hybrid functional (HSE) are used for DFT calculations, with plane-wave basis.
The cut-off energy was 60 Ry (1 Ry = 13.60 eV).
The wave function can be solved non-consistently based on the output potential field file.
The energy band structure is obtained with calculate energy eigenvalues along a path between high symmetry points in a K space.
The total density of states (DOS) and the projected DOS according to atoms and orbital, are obtained by select 10 x 10 x 10 K points through a Monkhorst-Pack method.
The density of States is interpolated to smooth the curve.
\par
\begin{figure}[!t]
	\centering{\includegraphics[width=8.5cm]{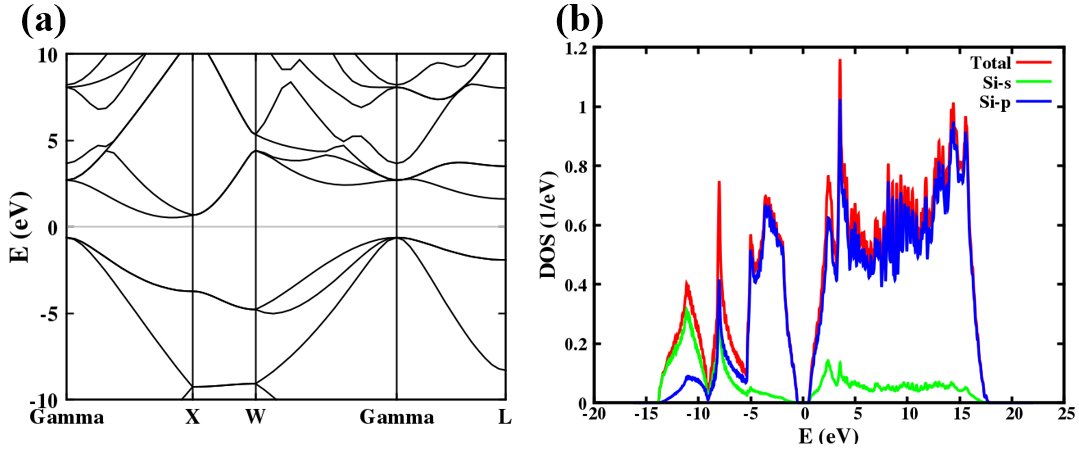}}
	\caption{(a) Band Structure and (b) total DOS and projected DOS of the primitive cell in bulk silicon at the Fermi level $E_{F}=0$ }
	\label{Si_BandDOS}
\end{figure}
\begin{figure}[!t]
	\centering{\includegraphics[width=8.5cm]{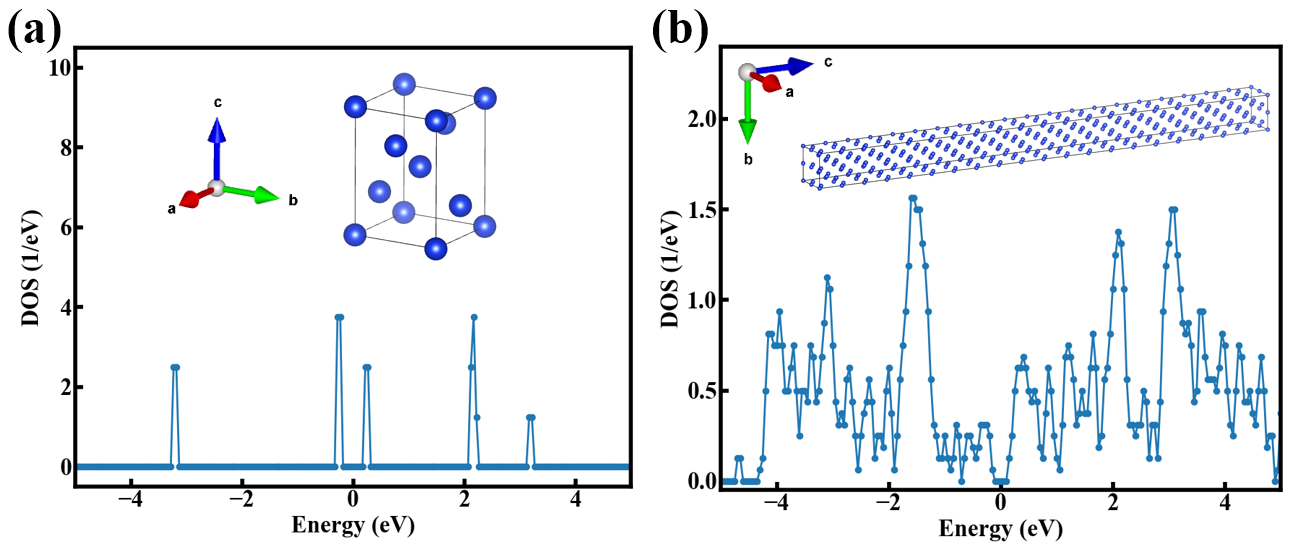}}
	\caption{(a) The density of states at the $\Gamma$ point of the unit cell along the [1 1 0], [1 -1 0], and [0 0 1]directions; (b) the density of states at the $\Gamma$ point of the supercell expanded by 2 x 2 x 20 times along axis directions. The volume of the Si supercell is 6405.5 $\AA^3$.}
	\label{Si_super_dos}
\end{figure}

The band structure and DOS of the bulk Si obtained by the above method are shown in Fig.\ref{Si_BandDOS}.
Because the NAMD module can only calculate the carrier dynamics of each state at $\Gamma$ point, it is necessary to expand the unit cell to fold the Brillouin zone and form a quasi-continuous energy level distribution at $\Gamma$.
In PWmat, the relaxed primitive cell is used to intercept the unit cell in the directions of [1 1 0], [1 -1 0] and [0 0 1], as shown in Fig.\ref{Siprim}(b).
The supercell structure for NAMD calculation is obtained by expanding the unit cell in the directions of a b c axes by $2 \times 2 \times 20$ times, respectively.
Its volume is 6405.5 $\AA^3$, and its DOS is shown in Fig.\ref{Si_super_dos}.
It can be seen that in the case of a small cell, the extracted state shows a discrete energy distribution due to the discrete energy level at the $\Gamma$ point.
After adopting a large supercell, the state at $\Gamma$ point via band folding shows a quasi-continuous distribution, and the density distribution of the state is closer to that calculated using multi k points as desired.
Then, the thermal process of electrons near the conduction band edge will be calculated based on the supercell.
\par
\subsubsection{The carrier energy relaxation in silicon}
\begin{figure}[!t]
	\centering{\includegraphics[width=8.5cm]{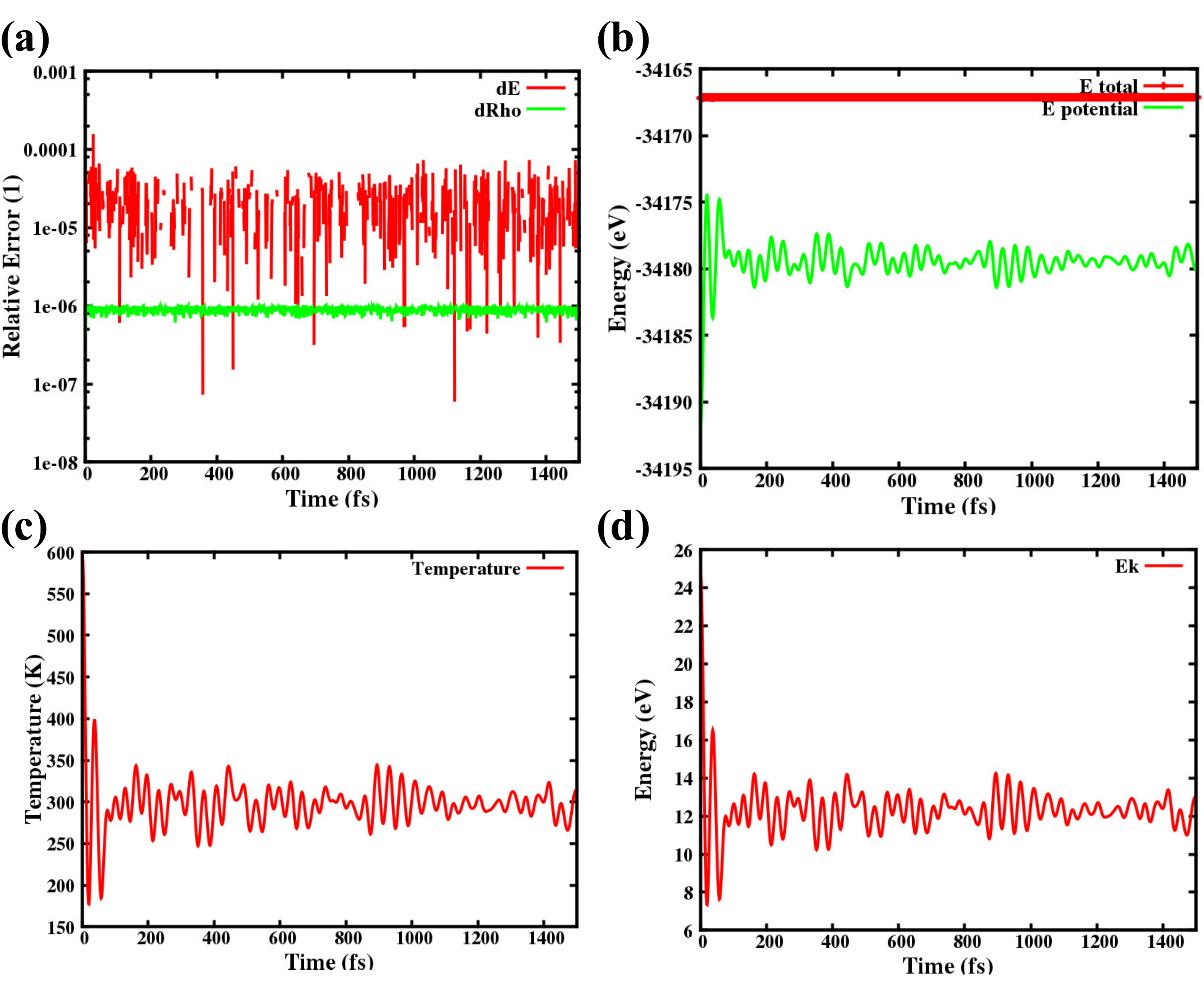}}
	\caption{BOMD calculation of Si super cell with 1500 fs (a) total energy and charge density relative error; (b) the change of total energy and potential energy with time; (c) temperature and (d) kinetic energy change with time.}
	\label{Si_BOMD}
\end{figure}

The total number of atoms in the Si supercell is 320, the number of valence electrons is 1280.
The BOMD simulation settings are as follows, the number of molecular dynamics steps is 1500, the step size is 1 fs, the total number of energy bands is 1600, and the exchange correlation functional is parameterized by PBE under GGA.
The relative error of energy convergence is less than 1 $\times$ $10^{-8}$.
Since the energy convergence does not necessarily guarantee the convergence of charge density, the relative error of electron density convergence is set to be less than 1 $\times$ $10^{-6}$.
If one of them reaches the criteria, the convergence is achieved.
The result is shown in Fig.\ref{Si_BOMD}(a).
To reduce the tiny vibration of the eigenlevels, a more stringent wavefunction convergence criteria of 1 x $10^{-6}$ is set.
During BOMD calculation, 101 energy levels at and above the conduction band edge are required to be output for subsequent NAMD calculation.
The total energy and potential energy variations are calculated at 300 K as shown in Fig.\ref{Si_BOMD}(b), the atomic temperature and kinetic energy changes are shown in Fig.\ref{Si_BOMD}(c) and (d).
As the result shows that total energy always keeps the same under the convergence condition, the kinetic energy and the potential energy convert into each other.
The initial assumption is set with twice of the kinetic energy.
After about 100 fs of intensive oscillation, the kinetic energy and potential energy reach equipartition and the total energy is around $320 \times \frac{3}{2}k_BT = 12.43$ $eV$.
Even after 200 fs, temperature can keep around 300K, while lattice vibration tends to be fairly regular.
Therefore, in the following NAMD calculations, the first 200 fs with intensive vibration with 200 fs is removed.
\par
\begin{figure}[!t]
	\centering{\includegraphics[width=8.5cm]{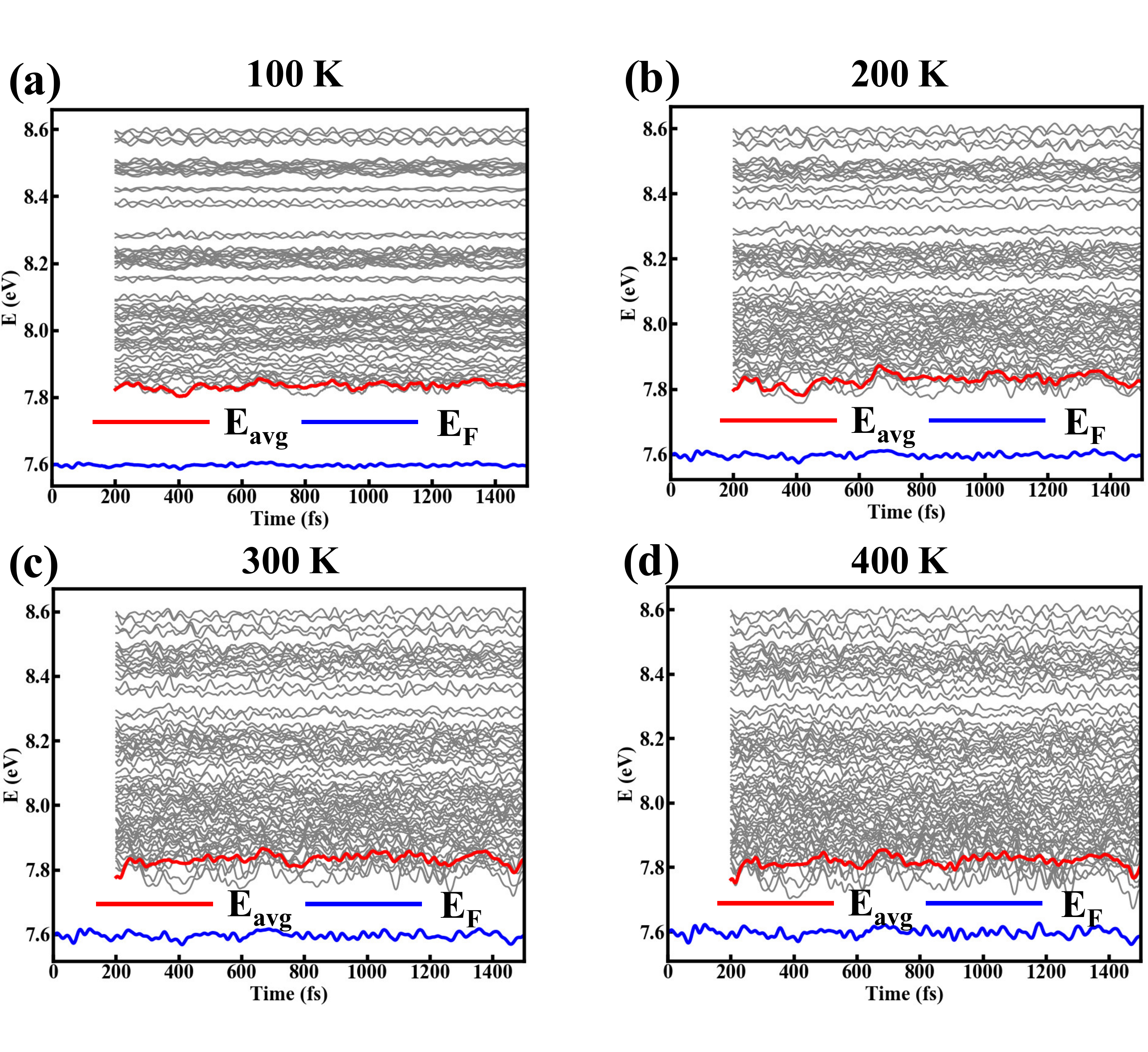}}
	\caption{Thermalization process of Si supercell at (a) 100K, (b) 200K, (c) 300K, (d) 400K in 1500 fs calculated by NAMD. The gray lines in the figure are the eigenvalues of different energy levels at and above the conduction band edge of Si. The red line is the average energy $E_{avg}$of each level according to the occupation weight, and the blue line is the Fermi level $E_F$.}
	\label{Si_NAMD_temp}
\end{figure}

In order to investigate the different initial atomic vibrations, the initial time interval of 20 fs from 200 fs to 400 fs is set for NAMD calculations.
The initial electron distribution in energy is set at the lowest energy level, which means the occupancy on the conduction band miminum (CBM) is 1, the occupancy on the other energy levels is 0.
Note that the carrier relaxation does not sensitively depend on decoherence time, in this calculation we set it as 80 fs~\cite{kang2019nonadiabatic}.
Fig.\ref{Si_NAMD_temp} shows the change of each eigenenergy with time in the specific window at different temperatures calculated by NAMD.
The red line and the blue line are the average energy of electrons and the Fermi level respectively.
The non-adiabatic window selects 101 states at conduction band above the band edge.
\par
At the low temperature (100K), due to the small lattice vibration and the small change in the potential field to the electron, the wave function and the eigenenergy level obtained by the solution change little with time, and the overlap between the energy states is small.
Because the distribution probability decays rapidly with the increase of energy, the electron occupation near the band edge is dominant, and the average energy of electrons changes from completely in the band edge at the beginin to slightly higher than the band edge after scattering.
If temperature further decreases to 0 K, all atoms are frozen and the potential field will not change with time.
So as the wave function and the energy level are obtained by stationary Schr$\ddot{o}$dinger equation.
Electrons are always in the initial energy level and will not scatter to other energy levels.
If the temperature is increased to 200 K, 300 K and 400 K (as shown in Fig.\ref{Si_NAMD_temp}(b)-(d)), it can be seen that with the increase of atomic vibration, the vibration of Fermi level and eigenlevel is intensified.
And the overlap between different eigenstates becomes greater, which intensifies the scattering of electrons between states. T
Then influence of introducing phonons on the electron dynamics process can be evaluated.
The average energy of electrons is also higher than that at low temperature, because the probability of electron distribution at high temperature decays less with energy and the energy distribution of electrons is broadened.
\par
\begin{figure}[!t]
	\centering{\includegraphics[width=8.5cm]{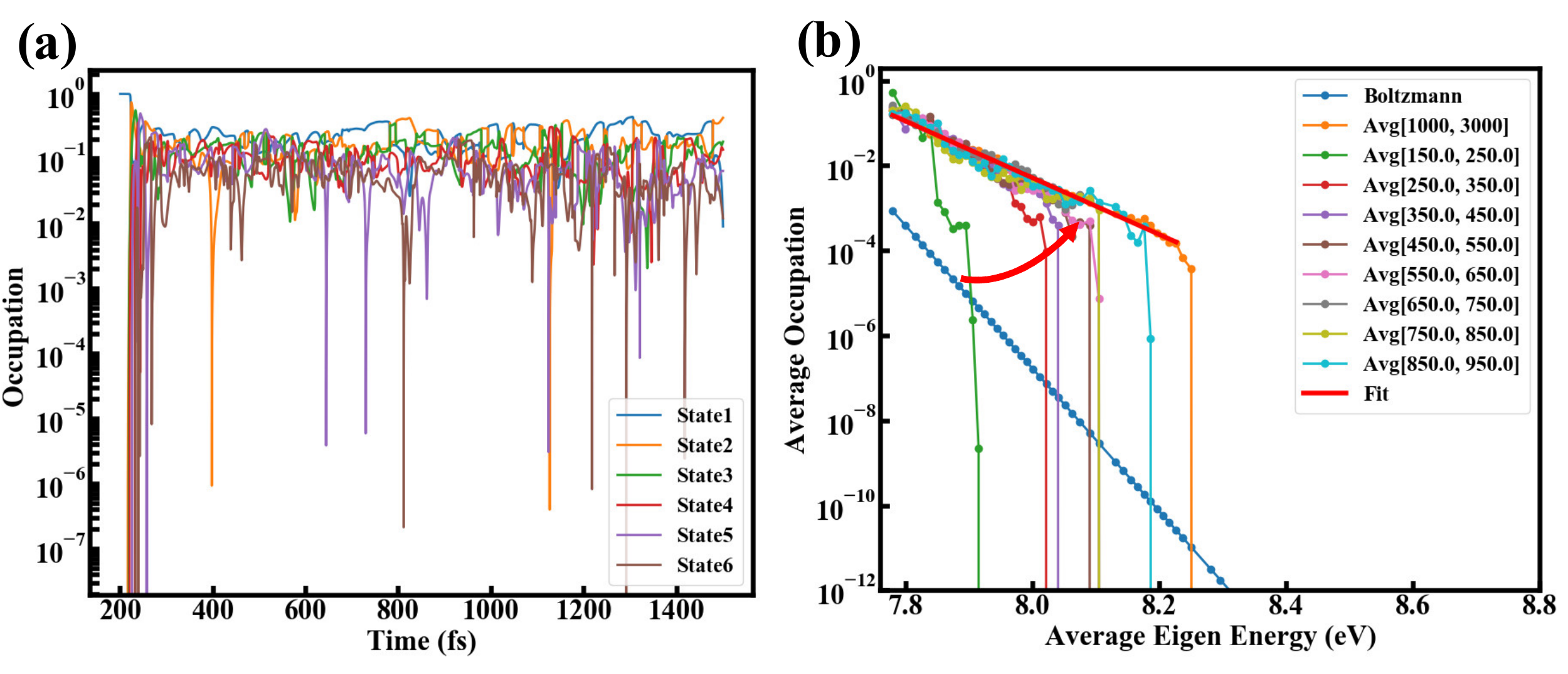}}
	\centering{\includegraphics[width=8.5cm]{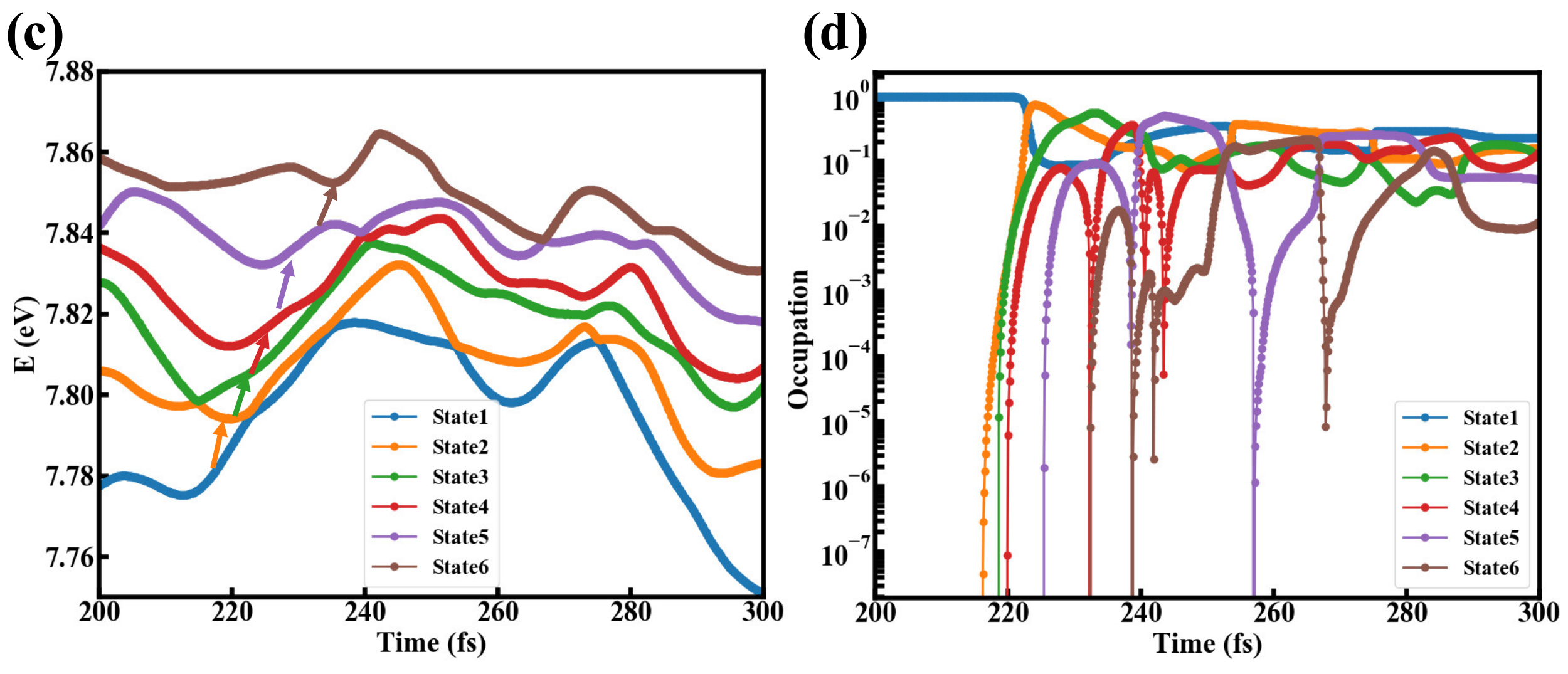}}
	\caption{The NAMD process of Si supercell from 200 fs to 1500 fs at 300 K, (a) the change of occupancy probability over time at six energy levels near the band edge; (b) the change of average occupancy rate at different energy levels within 100 fs; Changes in (c) intrinsic energy and (d) corresponding occupancy probability of the six energy levels near the 200 to 300 fs band edge.}
	\label{Si_NAMD300K_Occ}
\end{figure}

To address the thermal process at 300 K more clearly, we then focus on the specific occupancy probabilities of the lowest 6 energy levels near the band edge changes with time as shown in Fig.\ref{Si_NAMD300K_Occ}(a).
Initially, electron only occupy the first energy level of the conduction band, i.e., CBM.
The occupancy probability of higher energy levels increase from 0 due to the in-elastic scattering process.
Since the lattice is always in vibration and the occupation probability of each energy level is constantly oscillating, we extract the change of the average occupation with the average intrinsic energy level in a period, as shown in Fig.\ref{Si_NAMD300K_Occ}(b).
From the initial time, the electron occupation gradually transits from some lower energy levels to higher energy levels, and the occupation probability of each energy level gradually tends to be balanced.
After a sufficiently long time (more than 1000 fs at 300 K), the occupation probability reaches an equilibrium distribution that decreases exponentially with increasing mean eigenenergy.
Fig.\ref{Si_NAMD300K_Occ}(c) and (d) illustrate the corresponding relationship between the the intrinsic energy level oscillation and the occupation probability variation within 200 to 300 fs.
When the wave function of each energy state overlaps with each other, the intrinsic energy levels approach each other, and the electron can jump from the lower energy level to the higher energy level, so that the occupation probability of the higher energy level increases gradually.
Finally, the dynamic balance of scattering from low energy level to high energy level and from high energy level to low energy level is achieved.
In terms of energy, the average energy of electrons increases due to the lattice vibration, and the occupation of different energy levels shows a gradual thermalization process, i.e., scattering cold carriers to high-energy carriers.
Note that the BOMD calculation determines how the intrinsic energy level changes, and the scattering of carriers depends on the specific energy level oscillation.
The scattering process of different initial processes is different, so a series of different initial times are set, and the final statistical thermalization process gives the characteristics of the system.
\par

\subsubsection{The Carrier thermalization factor and scattering rate in silicon}
\begin{figure}[!t]
	\centering{\includegraphics[width=8.5cm]{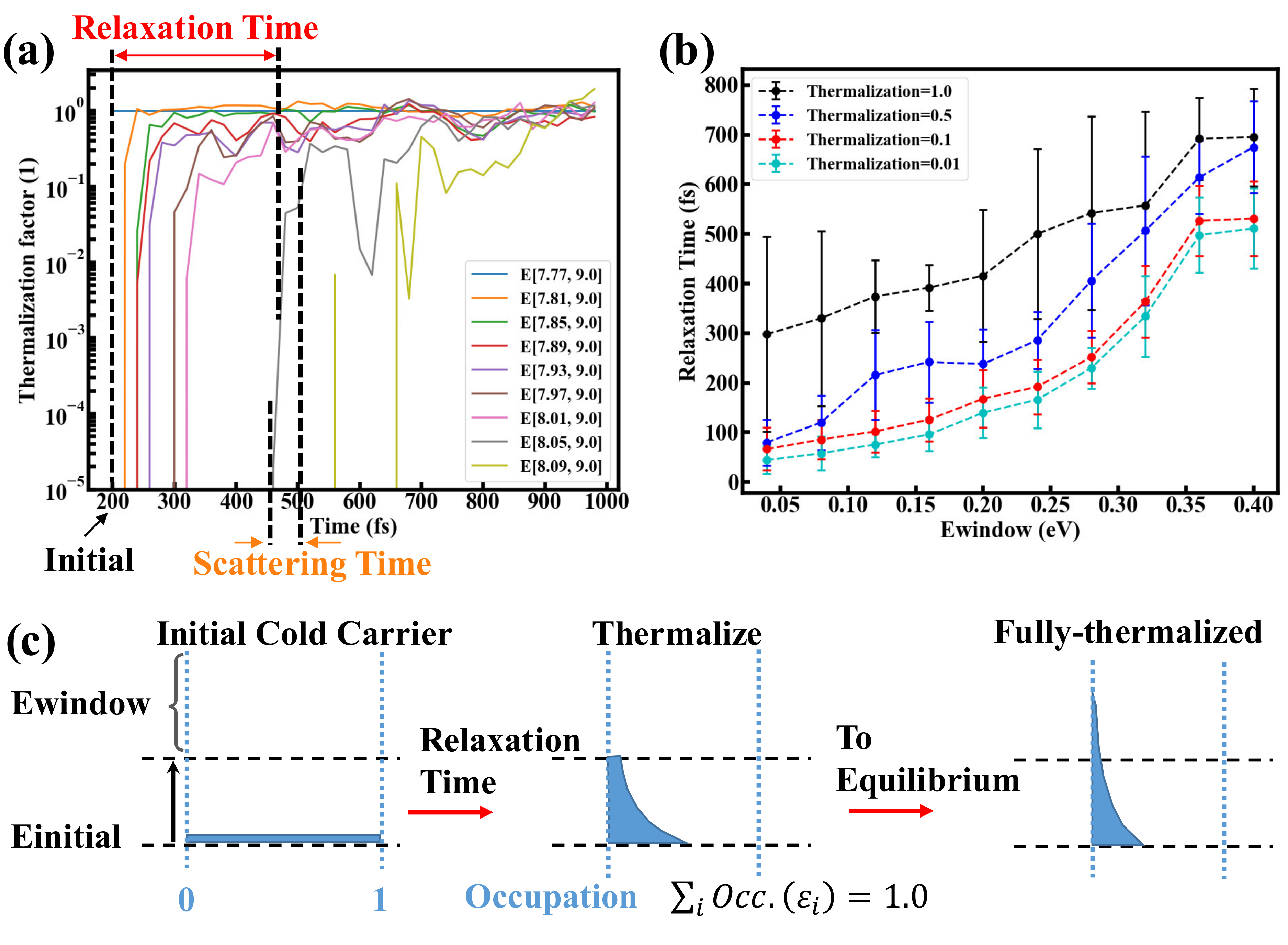}}
	\caption{Si supercell at 300 K (a) Thermalization factor versus time for each energy window from 200 fs to 1000 fs, (B) Relaxation time extracted from initial time to different thermalization levels, NAMD processes with an initial time of 200 fs to 400 fs and a step of 20 fs are counted. (C) Schematic diagram of the occupation probability of cold carriers at each energy level as a function of the thermalization process.}
	\label{Si_thermalization}
\end{figure}

In order to quantitatively characterize the thermalization degree of the cold carrier in the thermalization process, thermalization factor at a time of carrier in high energy window can be expressed as the current
total occupation distribution within the energy window divided by the total occupation distribution after reaching thermal equilibrium, as following function,
\begin{equation}
	\alpha(t) = \frac{\sum_{\epsilon_i \in E_{window}}Occ.[\epsilon_i(t)]}{\sum_{\epsilon_i \in E_{window}}Occ.[\epsilon_i^{eq}]}
\end{equation}
where $\epsilon_i$ is the energy eigenvalue of each energy level, $Occ.[\epsilon_i(t)]$ is the occupation probability of a certain energy level at a certain time, and $Occ.[\epsilon_i^{eq}]$ is the occupation probability of the intrinsic energy level after reaching thermal equilibrium. Fig.\ref{Si_thermalization} extracts the time line of the thermalization factor for each energy window from 200 fs to 1000 fs.
In the whole energy space (7.77 to 9.0 eV), the total occupation probability is normalized due to the conservation of particle number, and the thermalization factor always remains at 1.
For the energy window above the initial energy level, there is no occupation in the window at the beginning, and the thermalization factor is 0.
With the gradual thermalization of electrons caused by electron-phonon coupling, the low-energy electrons scatter into the high-energy window, and the thermalization factor increases.
After a sufficient time, the thermalization factor tends to 1.0 and oscillates around 1.0.
The reason for the oscillation is that the occupation probability of the high energy level may exceed the average occupation probability over a certain time or lower than the average occupation probability due to the change of the relative energy between the intrinsic energy levels with lattice vibration.
The average occupation is calculated at thermal equilibrium after 1000 fs.
Because the electron can only obtain a limited energy under the electron-phonon coupling, the thermalization process gradually takes place from the lower energy level to the higher energy level.
The thermalization in higher energy level starts after the electrons occupation reaching the its adjacent lower energy level.
So the relaxation time is defined as the total time from the initial time to the certain degree of thermalization in certain energy window.
A plot of the relaxation time to different degrees of thermalization ($\alpha$ = 0.01, 0.1, 0.5, and 1.0) as a function of the energy window, from the initial energy level, can be extracted as shown in Fig.\ref{Si_thermalization}(b).
The thermalization time, in general, increases as the energy window increases, since the relaxation of higher energy levels includes the relaxation time of lower energy windows.
In addition, the error bars in Fig.\ref{Si_thermalization}(b) show the standard deviation of the relaxation time.
\par
Thermalization to a higher thermalization degree (such as 1.0) has a larger fluctuation in the relaxation time than to a lower thermalization degree.
Because thermalization does not immediately or quickly reach thermal equilibrium near a thermalization degree of 1.0.
Instead, it gradually approaches thermal equilibrium through a series of scattering equilibria between high and low energy levels, such as some of the higher energy windows in Fig.\ref{Si_thermalization}(a).
In addition, an important phenomenon is that the time from the beginning of thermalization to thermalization to thermalization factor of 0.1 or 0.01 is quite fast.
The time scale of this part of the rising edge is in the order of tens of fs.
Therefore, most of the lower energy windows have been fully thermalized before lattice vibration exites carriers to higher energy windows.
For example, in Fig.\ref{Si_thermalization}(a), when the energy window of 7.97 to 9.0 eV starts thermalization, the thermalization of the lower energy window is already greater than 0.1.
Based on the above discussion, the physical image shown in Fig.\ref{Si_thermalization}(c) is established.
The cold carriers are initially set at the lowest energy level, and after a period of relaxation time, the low energy range is fully thermalized, while the high energy part is not thermalized effectively.
When the thermalization begins in a energy window, the carriers under this window can reach the high energy range through faster scattering, thus participating in the energy relaxation process step by step.
Finally, after a sufficient time of mutual scattering between high and low energy levels, the thermal equilibrium distribution is achieved.
\par
\begin{figure}[!t]
	\centering{\includegraphics[width=8.5cm]{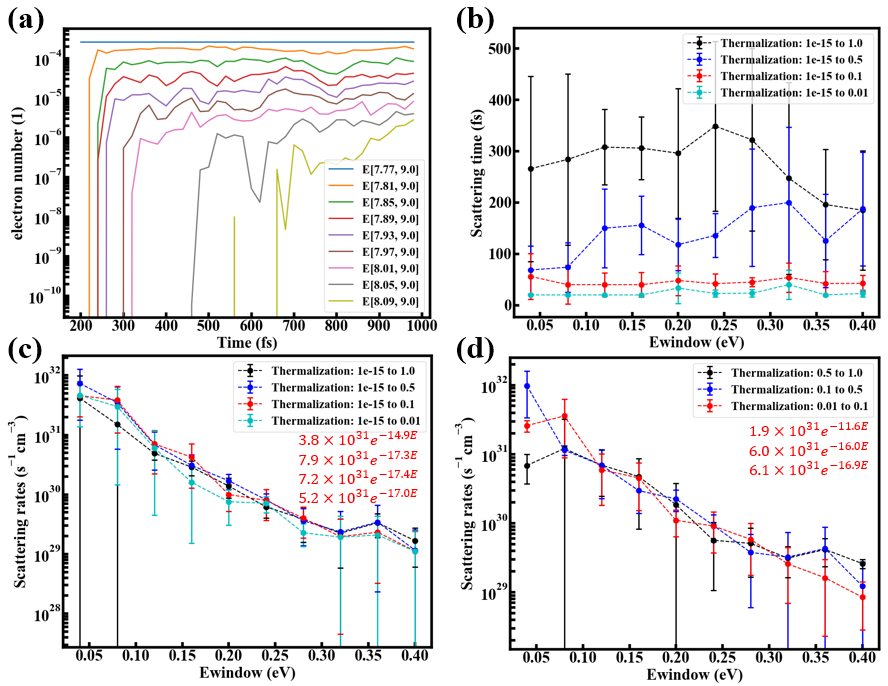}}
	\caption{Bulk Si, (a) the number of electrons in different energy Windows varies with time, the initial time is 200 fs, and (b) the scattering time from the beginning of thermalization to different thermalization degrees. Electron scattering (c) from initial thermalization to different degrees of thermalization and (d) partial thermalization. NAMD processes with initial time from 200 fs to 400 fs and steps of 20 fs were counted.}
	\label{Si_Scattering}
\end{figure}
In order to calculate the scattering rate of electrons, the number of carriers in the energy window per unit volume is calculated at firstly.
The total occupancy can be obtained by summing the occupancy of each state in the specific energy window $E_{window}$.
Since the occupancy in the whole NAMD window is normalized to $\sum_i Occ.(\epsilon_i(t)) = 1.0$, the number of electrons ($N_E(t)$) in $E_{window}$ is,
\begin{equation}
	N_E(t)=\frac{\sum_{\epsilon_i \in E_{window}}Occ.[\epsilon_i(t)]}{1.0} \int_{\epsilon_0}^{+\infty}g(\epsilon)f(\epsilon)d\epsilon
\end{equation}
where $\epsilon_0$ is the eigenenergy of the lowest energy level of the NAMD window.
For bulk Si, it is assumed that the Fermi level is at the edge of the band in the case of heavy doping, and the number of electrons in each energy window above CBM can be calculated according to the DOS of the primitive cell, as shown in Fig.\ref{Si_Scattering}(a).
It can be seen that the number of electrons is conserved in the whole window ranging from 7.77 eV to 9.0 eV).
The relaxation time of higher energy Windows is longer, and the number of electrons after reaching equilibrium also decreases exponentially.
This feature is consistent with the thermalization process illustrated in Fig.\ref{Si_thermalization}(a) and (b).
On the other hand, the extracted scattering time is shown in Fig.\ref{Si_Scattering}(b).
The scattering time is defined as the time from the starting of thermalization to a certain degree of thermalization in a specific energy window, that is, the rising edge time of the number of electrons in such window.
Note that the scattering time and relaxation time are different. The scattering time exhibits no obvious dependence on the energy position.
Instead, the scattering process depends on the interaction between several neighboring energy levels, rather than on the initial energy position.
The scattering time for thermalization to a thermalization factor of 0.1 is around 30 to 50 fs, while the scattering time for thermalization to a thermalization factor of 0.5 is 100 to 200 fs.
So electron in the high-energy window portion has a shorter scattering time than its relaxation time.
Moreover, like the relaxation time, the fluctuation of the scattering time for thermalization to a higher degree ($\alpha \geqslant 0.5$) is also larger.
Furthermore, the scattering rate ($S_E$) in the energy window can be calculated by the change in the number of electrons in the window,
\begin{equation}
	S_E=\frac{\Delta N_E}{V_C \Delta t}
\end{equation}
where $V_C$ is the volume of the cell, $\delta N_E$ is the change in the number of electrons under different degrees of thermalization, $\Delta t$ is the scattering time under different degrees of thermalization, $S_E$ represents the number of electron scattered into the high-energy window per unit volume per unit time.

The extracted scattering rates from the starting of thermalization to partial thermalization ($\alpha$= 0.01/0.1/0.5) and full thermalization ($\alpha$= 1.0) are shown in Fig.\ref{Si_Scattering} (c), and the scattering rates between different degrees of thermalization are shown in Fig.\ref{Si_Scattering}(d).
Some common characteristics can be found, the scattering rate decreases exponentially with the increase of the energy window.
Since there is no DOS truncation in the thermalization region.
The high energy part can be occupied by DOS, so the scattering rate will not be truncated.
The occupation probability of the high energy part decreases exponentially with the energy, because the probability of scattering of electrons to the high energy window also decreases exponentially.
According to the fitting data, the $S_E$ of scattering to 1.0 is slightly smaller than scattering to thermalization of 0.5 and 0.1.
Because in the transition process from partial thermalization to complete thermalization, the occupation of low-energy states decreases while the occupation of high-energy states increases.
The scattering rate from high-energy to low-energy states also increases and the net scattering rate in the energy window decreases.
However, the scattering rate of low thermalization degree such as (0.5 and 0.1) does not change much.
Because the occupation of low energy state is still high, and the occupation of high energy state is still low, the net scattering rate is mainly dominated by the scattering from low energy to high energy window.
Similarly, the extraction of partial thermalization scattering rate also shows the same phenomenon, as shown in Fig.\ref{Si_Scattering}(d).
There is no distinct difference in scattering rates corresponding to rather lower thermalization degrees.
While for rather high thermalization degrees, the thermalization scattering rate decreases significantly as the degree of thermalization increase, e.g., from 0.5 to 1.0.
Based on the extracted scattering rate, the transport process of the cold source transistor with thermalization, and the influence of thermalization on the leakage current can be given semi-quantitatively, which will be compared with the thermalization influence of the metal contact layer.

\par
	\begin{figure}[!tb]
	\centering{\includegraphics[width=8.5cm]{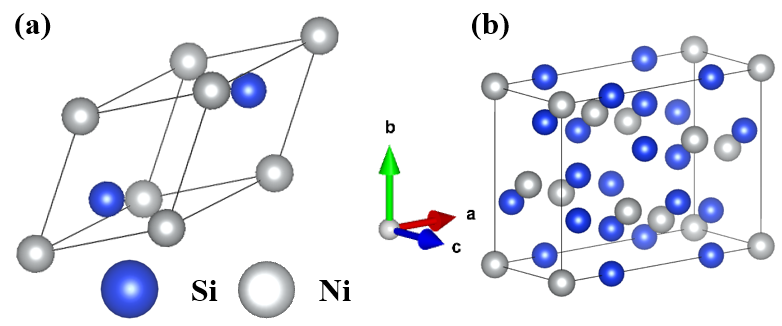}}
	\centering{\includegraphics[width=8.5cm]{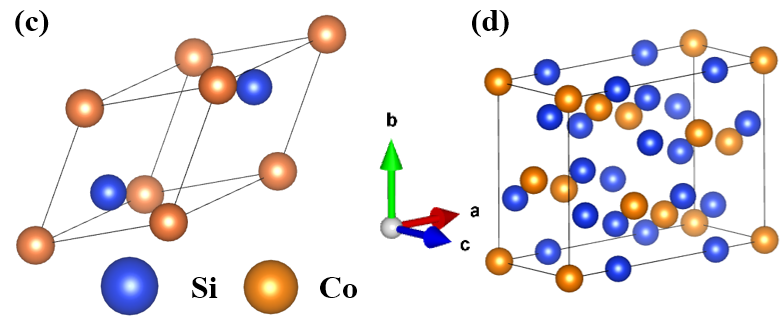}}
	\caption{(a) structure of NiSi$_{2}$  primitive cell with 3.829 \AA lattice constant; (b) NiSi$_{2}$  unit cell cut in [1 1 1], [1 1 -2], [1 -1 0] direction; the volume of primitive cell and unit cell is 39.7 $\AA^3$ and 238.3$\AA^3$, respectively; (c)structure of CoSi$_{2}$  primitive cell with 3.787 \AA lattice constant; (d) CoSi$_{2}$  unit cell cut along same direction; the volume of primitive cell and unit cell is 39.4 $\AA^3$ and 230.5$\AA^3$, respectively; In figure, a, b and c axis directions are [1 1 1], [1 1 -2], [1 -1 0], respectively. }
	\label{NiSi2Prim}
\end{figure}

	\begin{figure}[!tb]
	\centering{\includegraphics[width=8.5cm]{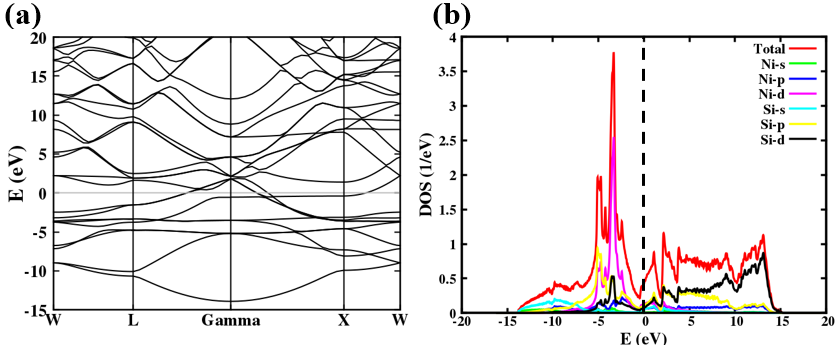}}
	\centering{\includegraphics[width=8.5cm]{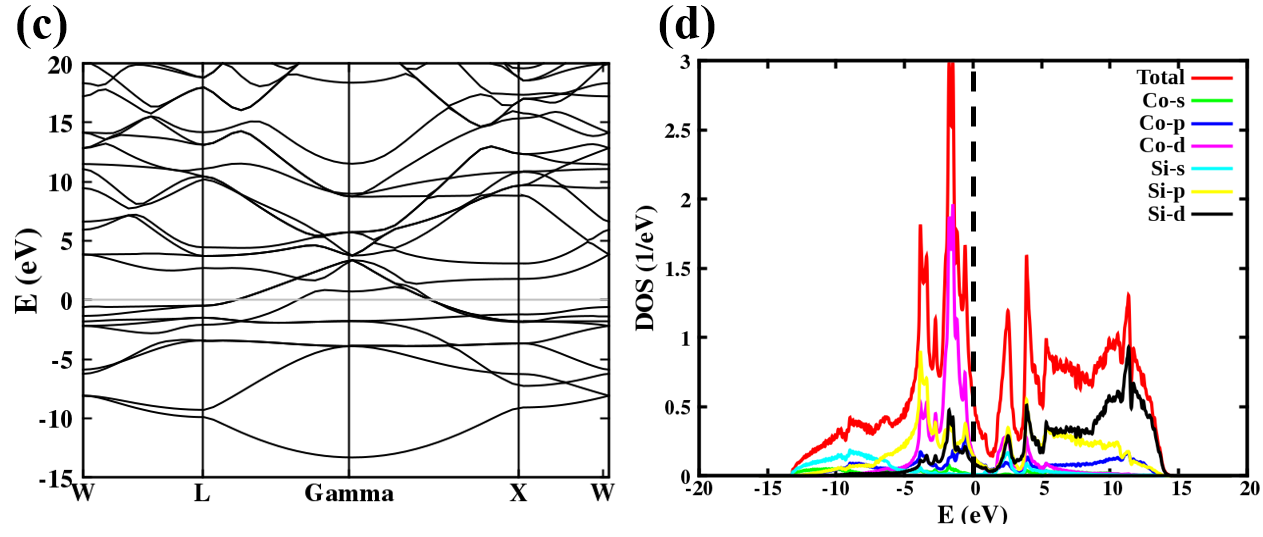}}
	\caption{(a) Band structure and (b) DOS of NiSi$_{2}$  primitive cell; (c) Band structure and (d) DOS of CoSi$_{2}$  primitive cell; the Fermi level is set to 0 eV.}
	\label{SilicideBandDOS}
\end{figure}

\subsection{The carrier thermalization process in metallic silicide}
\subsubsection{Crystalline structure of NiSi$_{2}$  and CoSi$_{2}$}
\begin{figure}[!t]
	\centering{\includegraphics[width=8.5cm]{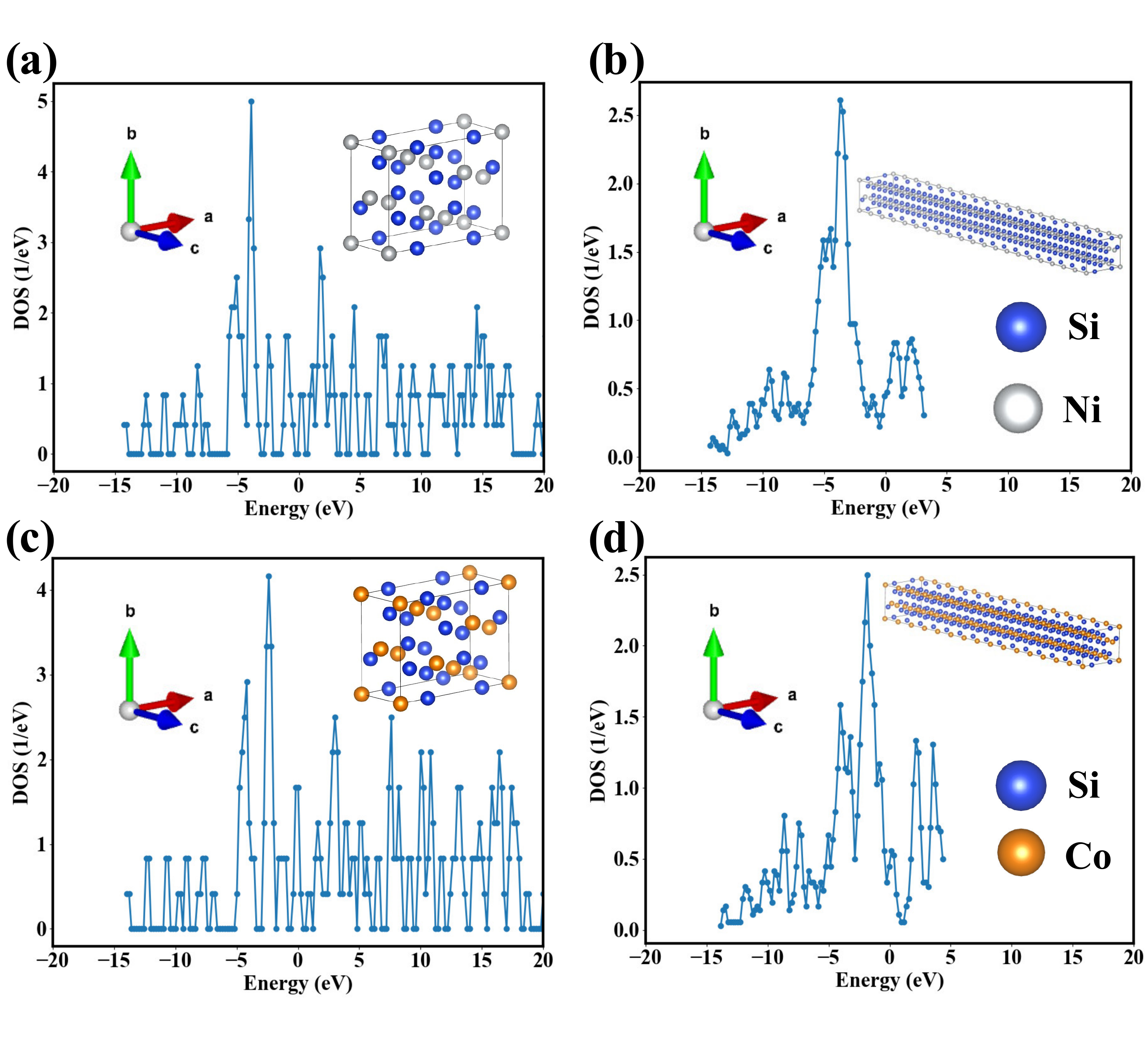}}
	\caption{the DOS of NiSi$_{2}$  at $\Gamma$ point (a) unit cell and (b) super cell which is expanded 2x2x20 along a, b and c axis; the DOS of CoSi$_{2}$  at $\Gamma$ point (a) unit cell and (b) super cell which is expanded 2x2x20 along a, b and c axis;}
	\label{NiCoSuper}
\end{figure}
\begin{figure}[!t]
	\centering{\includegraphics[width=8.5cm]{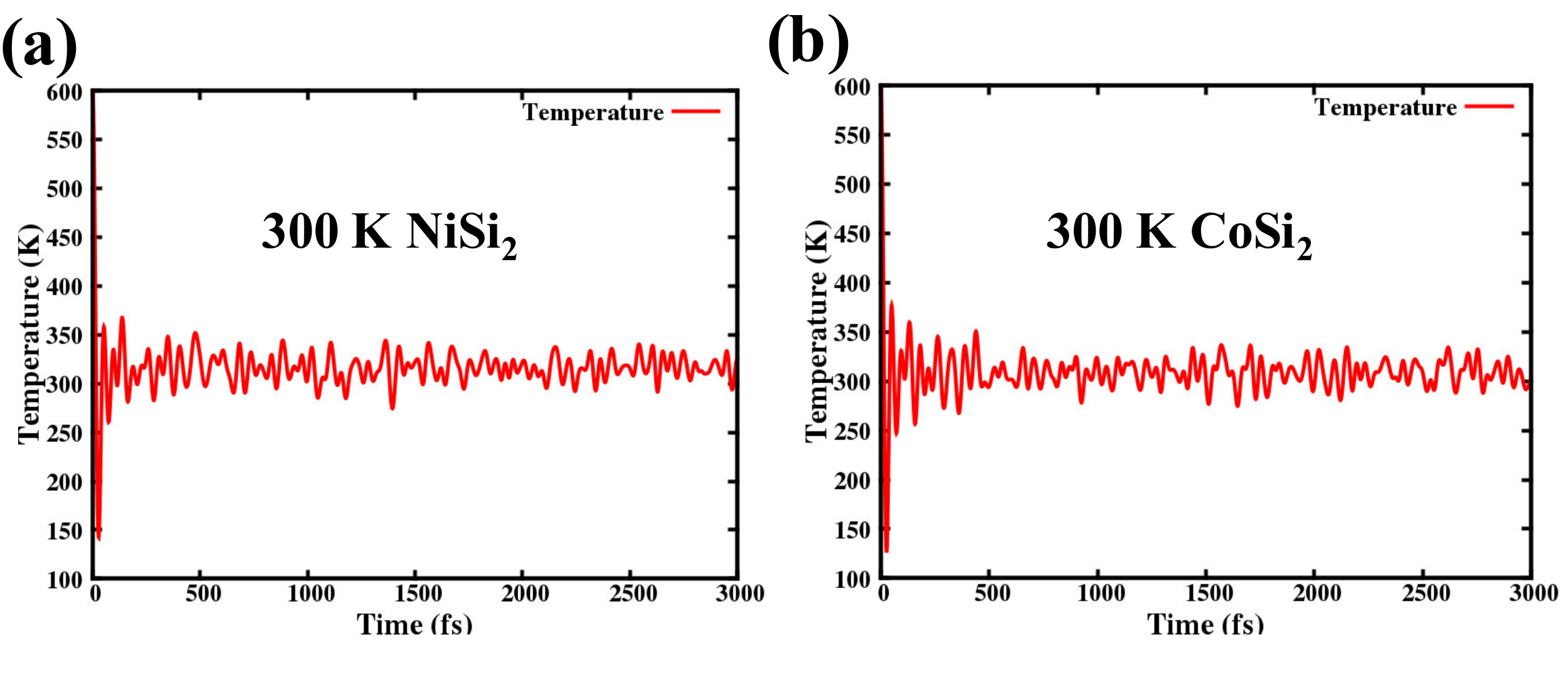}}
	\caption{Supercells of (a) NiSi$_{2}$  and (b) CoSi$_{2}$  as a function of temperature (lattice vibrational kinetic energy) over 3000 fs at a 300K.}
	\label{NiCoTemp}
\end{figure}

In the cold source FET proposal, a metallic insert layer is introduced to reduce the tunneling barrier as compared to regular tunneling FET~\cite{fliu2018,gan2020ted}.
Silicide/silicon interface can be formed in a highly controlled manner to ensure defect control and electrical properties of the metal/Silicon junction~\cite{tu1981apl}.
These advantages can be inherited to Silicon nanowire cold source FET~\cite{gan2021ted}.
Metallic silicide has a continuous state density near the Fermi level, so the energy-filtered carrier may be gradually thermalize to high energy, leading to leakage and $SS$ degradation.
It is necessary to investigate the influence of electron-phonon coupling on thermalization in the metallic silicide by NAMD.
In metallic silicides, NiSi$_{2}$  \cite{tung1983formation,tung1986schottky} and CoSi$_{2}$  \cite{tung1982growth} have the comparatively small lattice mismatch with Si.
Both of them can be grown on the (1, 1, 1) plane of Si, and NiSi$_{2}$  has the minimum lattice mismatch and CoSi$_{2}$  has a longer mean free path~\cite{rosencher1984transistor}.
Therefore, they are expected to be used as insert metallic layer in cold source transistor based on mainstream Si-process.
As the Fig.\ref{NiSi2Prim}(a) and (c) shown,  NiSi$_{2}$  and CoSi$_{2}$  adopt the fluorite structure, with similar lattice constants which are close to lattice constant of Si. Similar to the above calculation process of Si, the calculation of NAMD needs to expand the unit cell by intercepting the unit cell in the directions of [1 1 1], [1 1-2] and [1-1 0], as shown in Fig.\ref{NiSi2Prim}(b) and (d),  and then expand the cell in the [1-1 0] direction to obtain the supercell for NAMD calculation.
First, the band structure and density of States of the primitive cell are calculated.
After the primitive cell is relaxed, the atomic force is less than 0.01 eV/$\AA$.
The basis set is plane waves with a cut-off energy of 60 Ry.
The exchange correlation functional effect is calculated using the local density approximation (LDA).
The process of calculating energy band and state density is the same as that of Si in section~\ref{subsubsec:Si}.
The results of NiSi$_{2}$ and CoSi$_{2}$ are shown in Fig.\ref{SilicideBandDOS}, which presents similar result.
The main difference is the position of the Fermi level.
In CoSi$_{2}$, it has a lower Fermi level position because Co atom lack one $d-$ orbital electron than Ni.
For NiSi$_{2}$, the minimum of DOS is below Fermi level and DOS near Fermi level increases with energy, while for CoSi2, the minimum of DOS is above Fermi level and DOS near Fermi level decreases with energy.
The density of States at $\Gamma$ point calculated by the unit cell of Fig.\ref{NiSi2Prim}(b) and (d) shows a discrete distribution in Fig.11 (a) and (c).
Based on the NiSi$_{2}$ and CoSi$_{2}$ unit cells, the cells were expanded by 20 times along the [1 -1 0] (c axis direction) perpendicular to the (1 1 1) plane, and the supercells and the extracted DOS at the $\Gamma$ point are shown in Fig.\ref{NiCoSuper}(b) and (d).
It can be seen that the density of States of the metal silicide itself is relatively high, it can be better reversed after cell expansion.
Compared to Si which has a smaller density of states and uses a very large supercell, metal silicides can form a quasi-continuous DOS spectrum with smaller supercells (3574 $\AA^3$ for NiSi$_{2}$  and 3457 $\AA^3$ for CoSi$_{2}$).
The total energy and temperature changes during the BOMD process of NiSi$_{2}$  and CoSi$_{2}$  are calculated under the supercell structure.
The temperature changes with time are presented in Fig.\ref{NiCoTemp}.
It fluctuates significantly within the initial 500 fs and then tends to 300K under equilibrium. Similarly, in order to eliminate the effect of the initial violent shock on the scattering, the initial time from 500 fs to 980 fs is selected, and 25 groups of NAMD processes are calculated.
The initial electron is set at the first energy level above the Fermi level, the occupancy rate of the remaining energy levels is 0, and the decoherence time is 80 fs.
\subsubsection{Carrier energy relaxation of NiSi$_{2}$  and CoSi$_{2}$}
\begin{figure}[!t]
	\centering{\includegraphics[width=8.5cm]{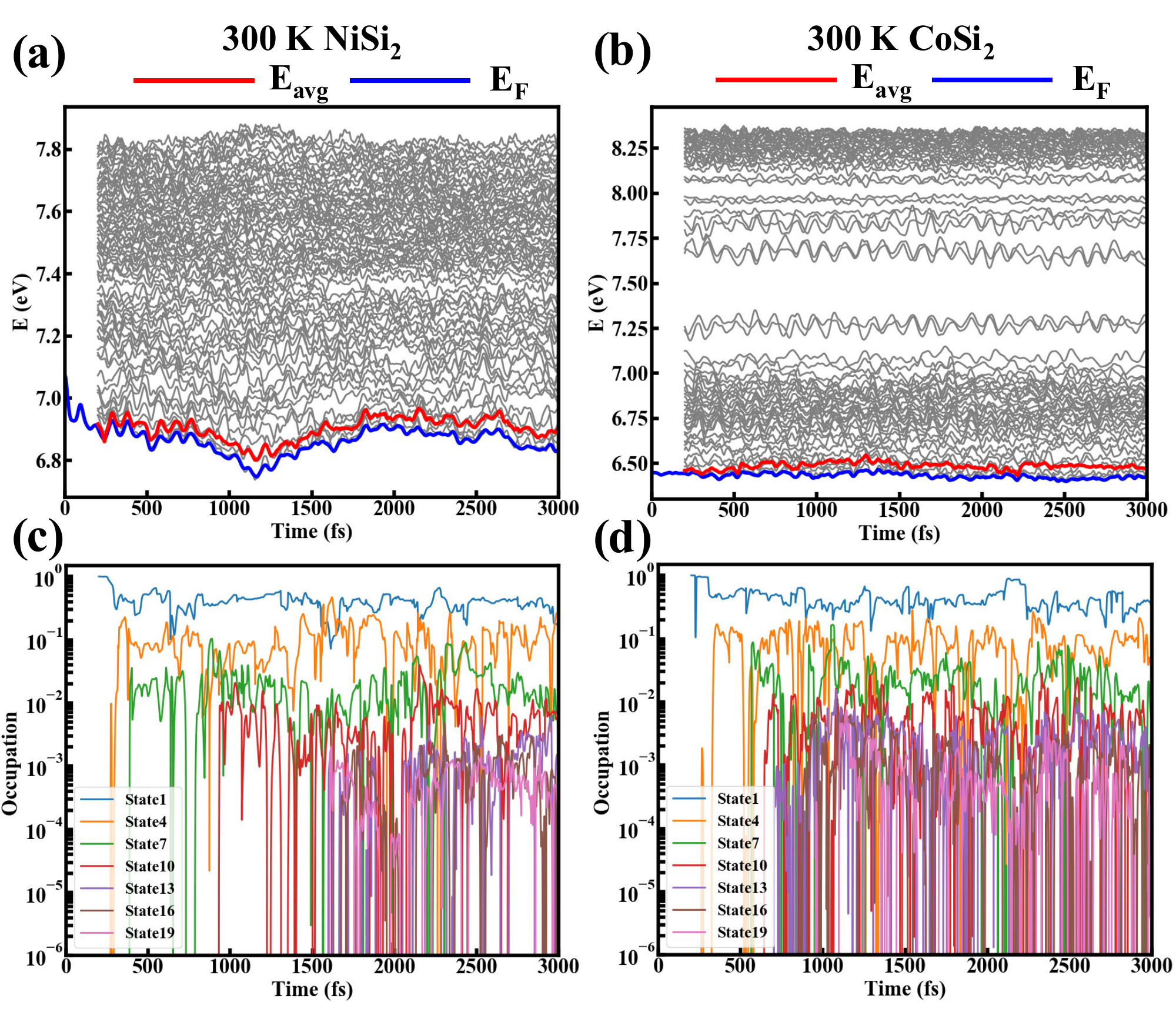}}
	\caption{Thermalization process of supercells of (a) NiSi$_{2}$  and (b) CoSi$_{2}$  at 300 K in 3000 fs. The blue line in the figure is the Fermi level $E_F$, and the gray lines are the eigenvalues of and above the different energy levels. The red line is the average energy $E_{avg}$ of each energy level according to the occupation weight.Time dependence of the occupation probability of energy levels above the Fermi level for the NAMD process from 500 fs initiation to 3000 fs for the supercells of (c) NiSi$_{2}$  and (d) CoSi$_{2}$, with state1 as the initial lowest energy level.}
	\label{NiCoNAMD}
\end{figure}

In the thermalization process of NiSi$_{2}$  and CoSi$_{2}$  at 300 K, the eigenenergies and average energies of the lowest 61 levels above the Fermi level are shown in Fig.\ref{NiCoNAMD}(a) and (b), and the initial time is 500 fs.
It can be seen that for NiSi$_{2}$  the DOS increases above the Fermi level (Fig.\ref{SilicideBandDOS}(b)), so that the intrinsic level above $E_F$ becomes progressively more dense, while for CoSi$_{2}$  there is a minimum above 1 eV above the Fermi level (Fig.\ref{SilicideBandDOS}(d)).
As a result, there is a relatively large gap above 0.75 eV for that intrinsic level above $E_F$.
In addition, the average energy has a small increase after time of thermalization.
The average energy reflects the occupation changes of the lower energy levels.
The occupation of higher energy levels cannot be reflected by the average energy but can be analyzed by the occupation probability of specific energy levels.
The specific occupation probability of the 1st to 19th States above the Fermi level varies with time, as shown in Fig.\ref{NiCoNAMD} (c) and (d).
The average energy increases as the lowest energy levels reach thermal equilibrium and begin to oscillate after being thermalized for hundreds of fs (State 1 to State 7).
Higher energy levels (State 10 and above) continue thermalization after 400 fs, but their equilibrium occupancy probability decays exponentially with energy.
The thermalization of the higher energy component makes little contribution to the average energy, and the average energy does not increase significantly further.
Therefore, it is difficult to reflect the thermalization effect of some energy levels through the change of average energy.
Alternatively, we extract the quantitative index of thermalization process directly through the occupation probability of each energy level.

\subsubsection{The carrier thermalization factor and scattering rate of NiSi$_{2}$  and CoSi$_{2}$}
\begin{figure}[!t]
	\centering{\includegraphics[width=8.5cm]{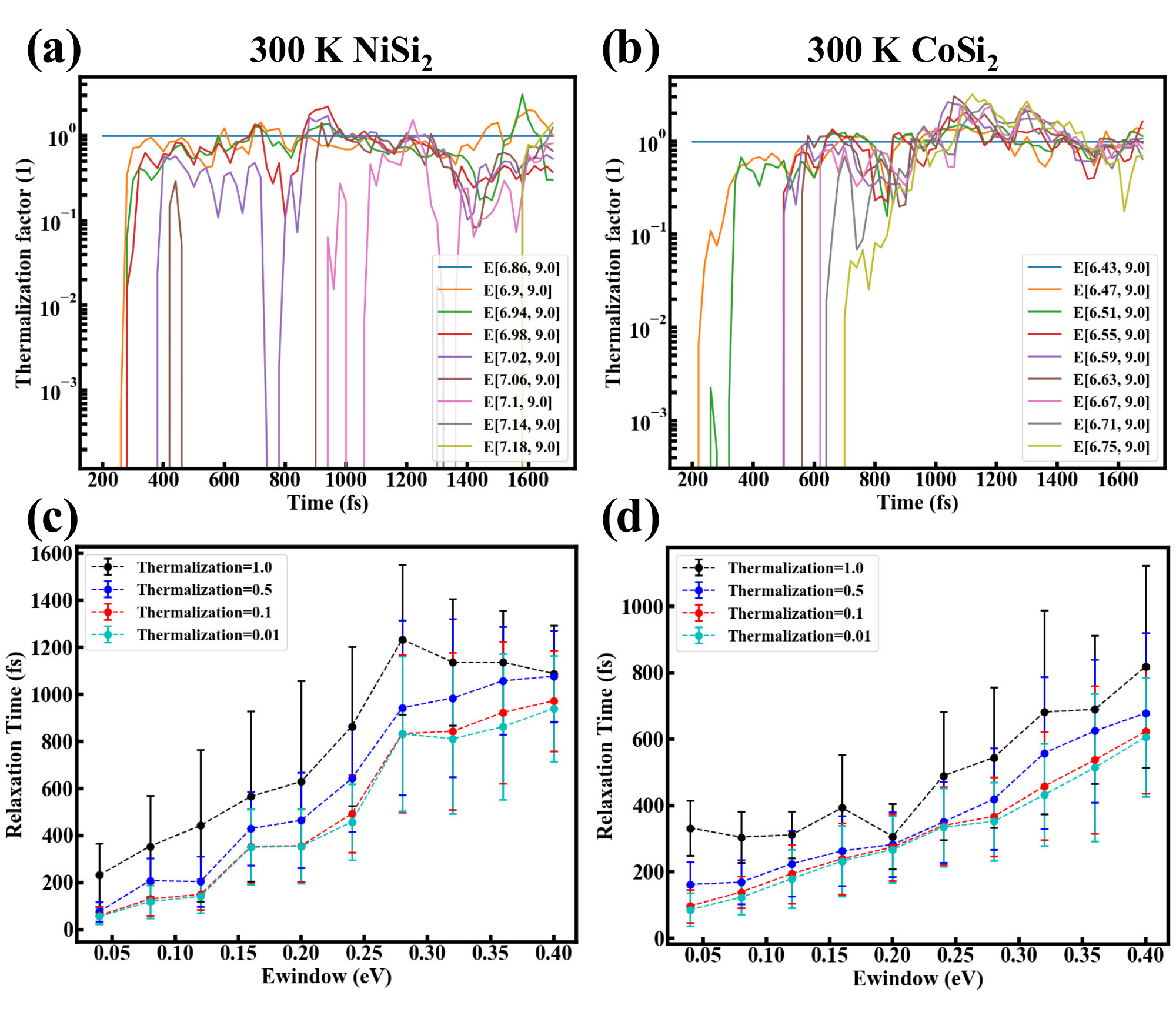}}
	\caption{Thermalization factor versus time at 300 K for supercells of (a) NiSi$_{2}$  and (b) CoSi$_{2}$  for each energy window. Relaxation times for supercells of (c) NiSi$_{2}$  and (d) CoSi$_{2}$  to reach different degrees of thermalization from the initial time. NAMD processes from 200 fs to 800 fs with a step of 20 fs are counted.}
	\label{NiCoThermalization}
\end{figure}

\begin{figure}[!b]
	\centering{\includegraphics[width=8.5cm]{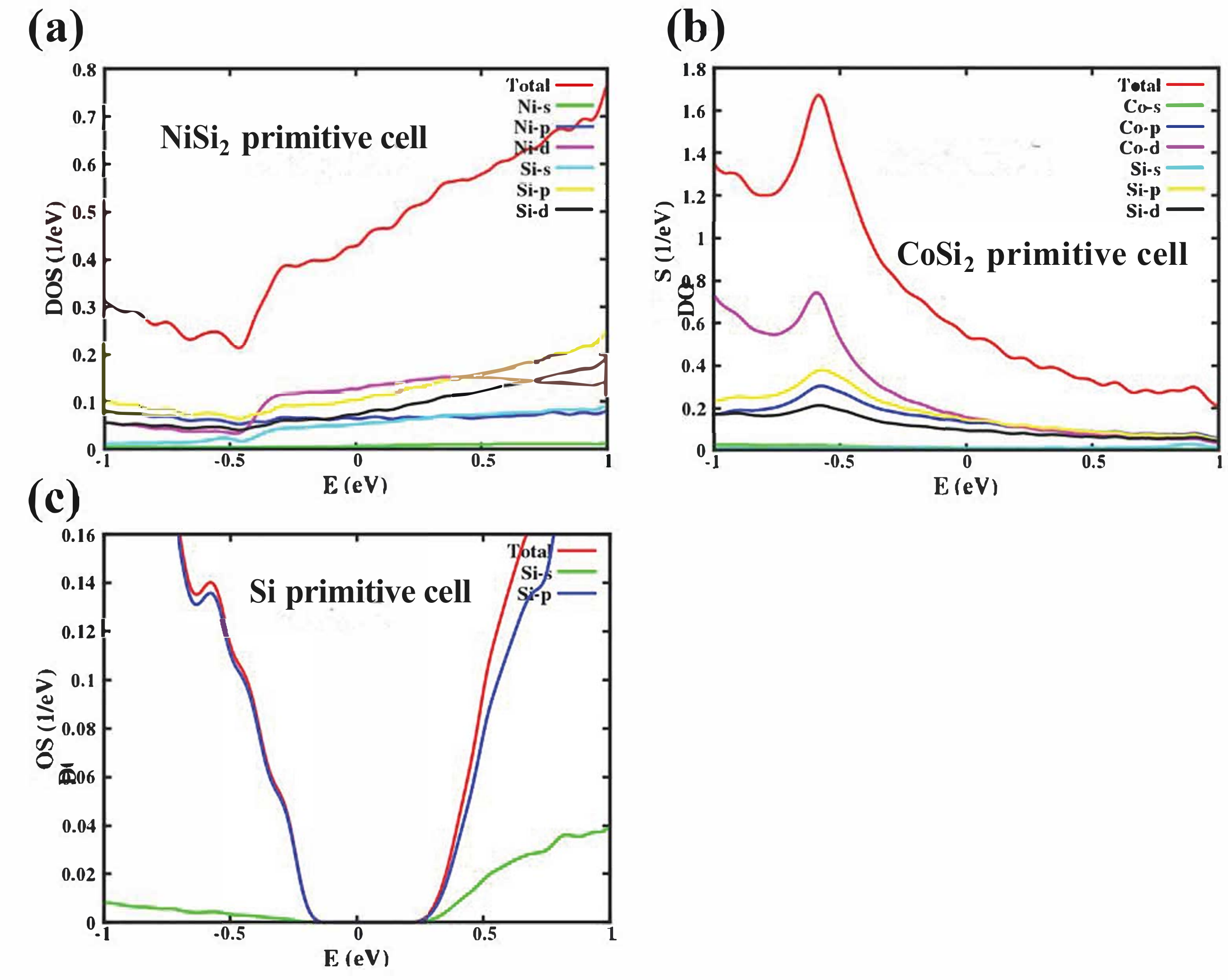}}
	\caption{DOS near the Fermi level for (a) NiSi$_{2}$, (b) CoSi$_{2}$  and (c)Si with primitive cell.}
	\label{NiCoSiDOS}
\end{figure}

\begin{figure}[!t]
	\centering{\includegraphics[width=8.5cm]{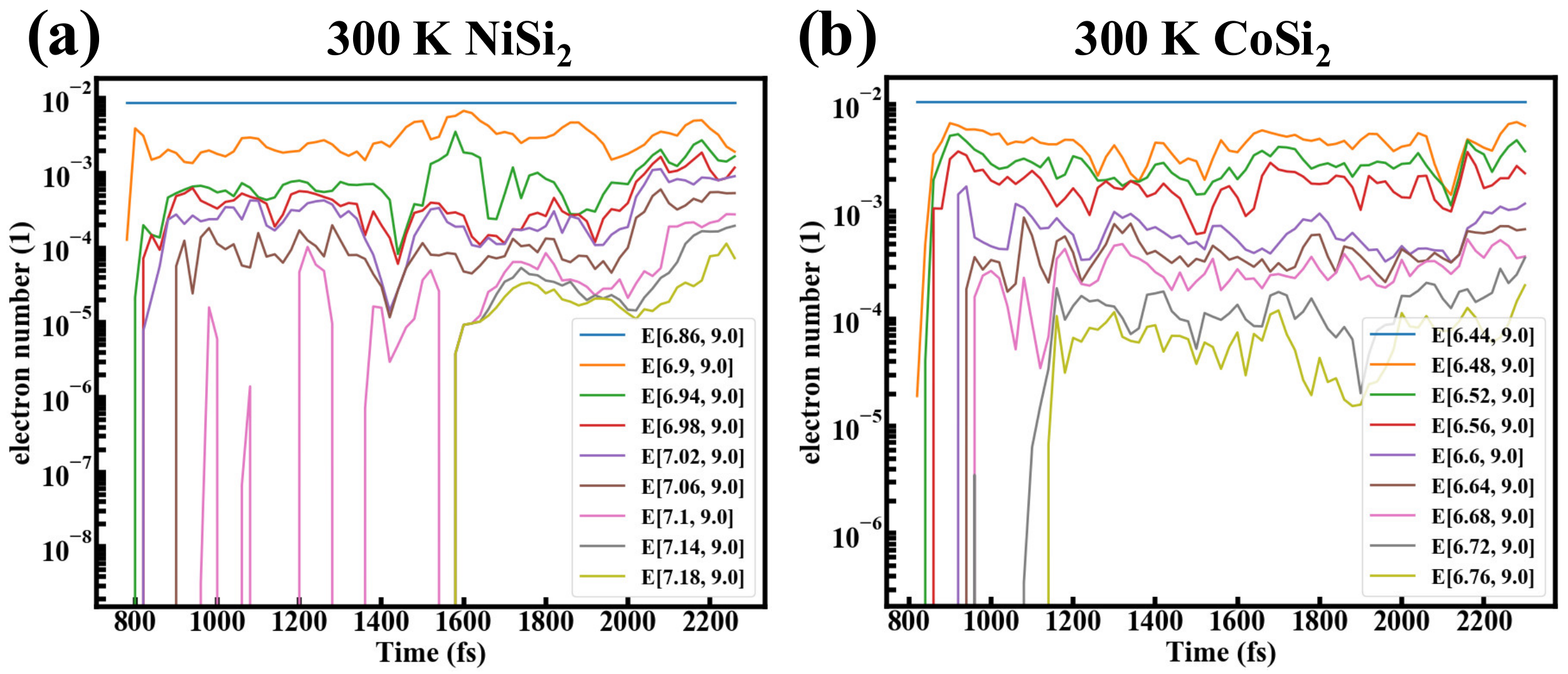}}
	\centering{\includegraphics[width=8.5cm]{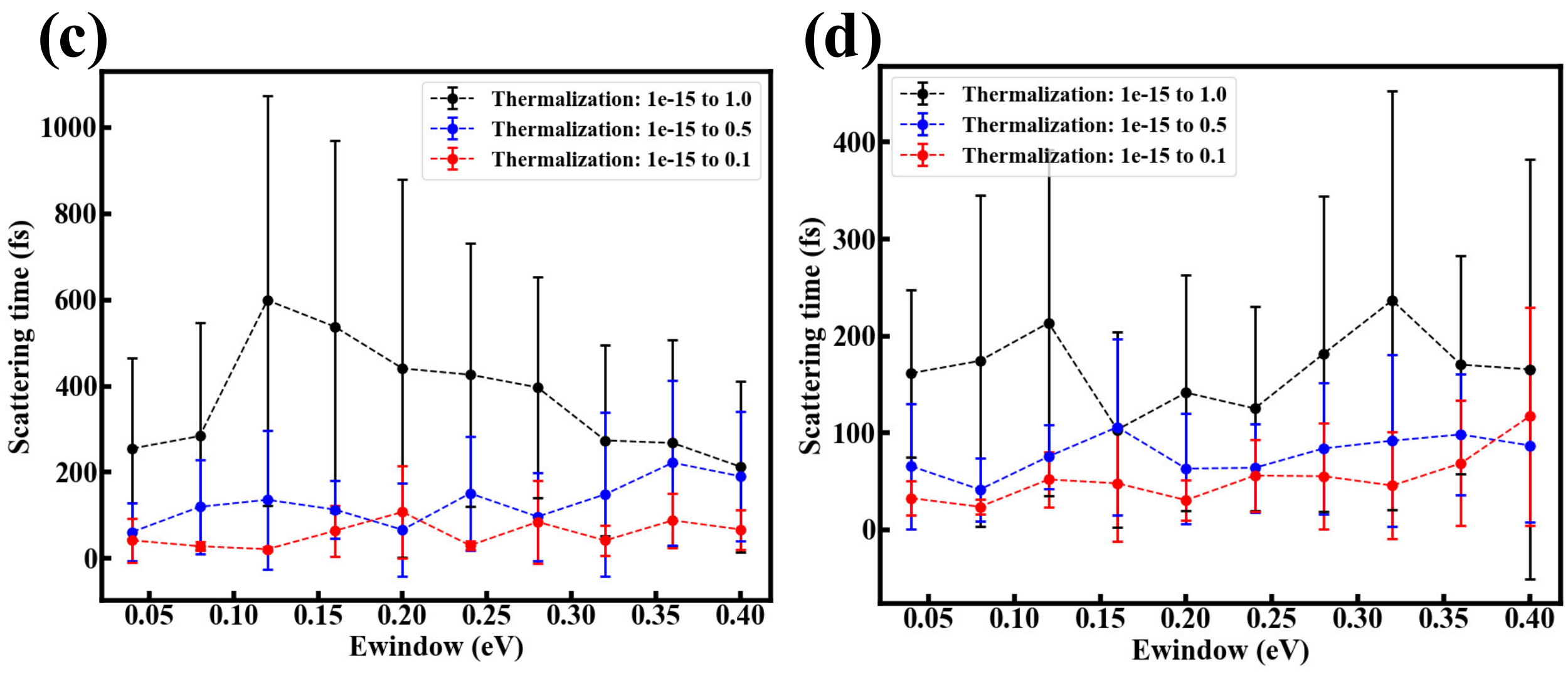}}
	\centering{\includegraphics[width=8.5cm]{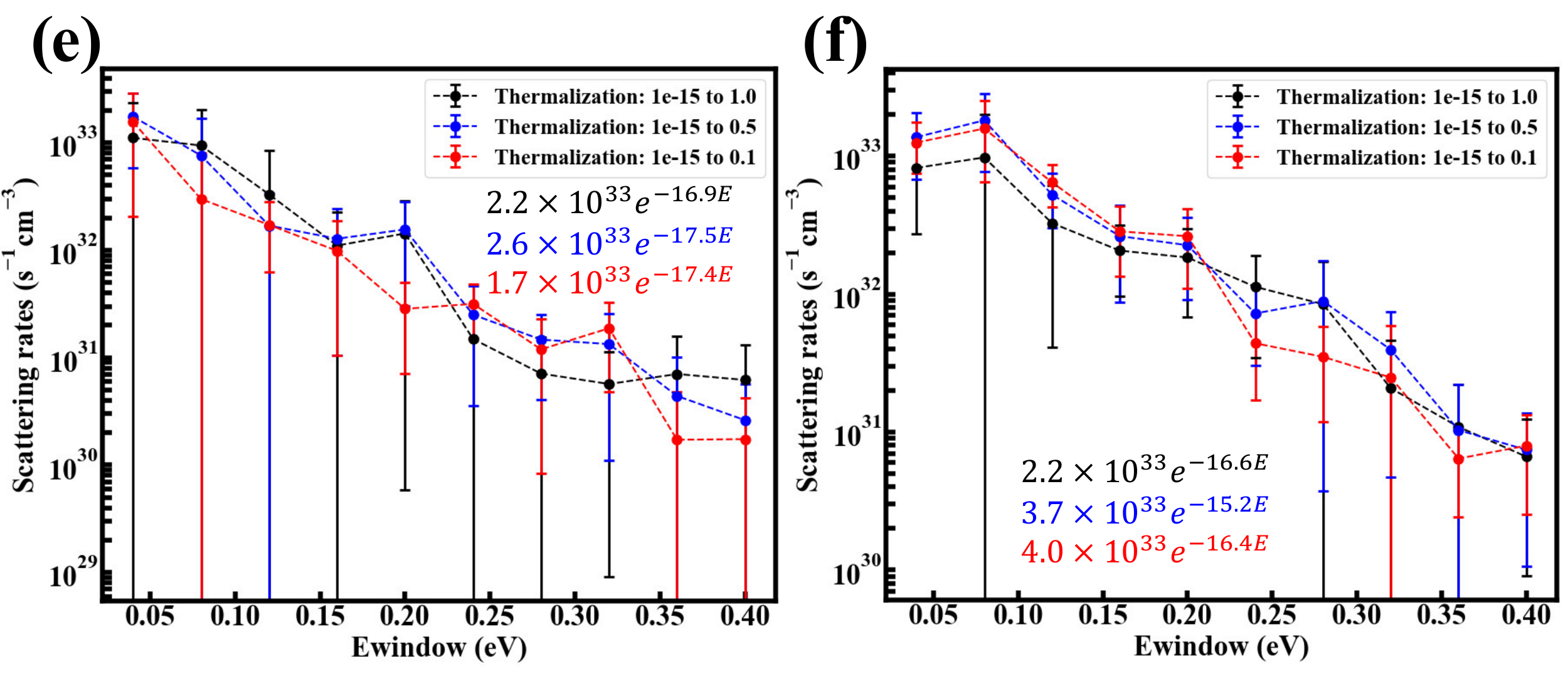}}
	\caption{Time dependence of the number of electrons in different energy windows for (a) NiSi$_{2}$  and (b) CoSi$_{2}$; Scattering time from the onset of thermalization to different degrees of thermalization for (c) NiSi$_{2}$  and (d) CoSi$_{2}$;  Scattering rate of (e) NiSi$_{2}$  and (f) CoSi$_{2}$ at different thermalization degrees;NAMD processes with steps of 20 fs from 500 fs to 980 fs were counted.}
	\label{NiCoScattering}
\end{figure}

Continually, the thermalization factors of NiSi$_{2}$  and CoSi$_{2}$  in each energy window can be extracted from the changes of their occupation probabilities with time, as shown in Fig.\ref{NiCoThermalization} (a) and (b).
The relaxation time from the initial time to a certain degree of thermalization is shown in Fig.\ref{NiCoThermalization} (c) and (d).
The thermalization process of NiSi$_{2}$  and CoSi$_{2}$  is similar to that of Silicon, i.e., the relaxation time depends on the position of the energy window from the initial energy level.
The partial thermalization relaxation time within 0.15 eV of the initial energy level is within 200 fs, while the relaxation time above 0.2 eV of the initial energy level is higher than 200 fs.
The relaxation time increases with the increase of energy window.
In addition, it can be seen from the relaxation time that after thermalization, the time difference between thermalization factor reaching 0.01 and 0.1 is very small.
From 0.1 to 0.5, the relaxation time difference is about 50 fs, and from 0.5 to 1.0, the relaxation time difference increases.
Therefore, the velocity between partial thermalization (thermalization factor from 0 to 0.5) is significantly higher than that between complete thermalization (thermalization factor from 0.5 to 1.0), indicating that the probability of high-energy window scattering back to low energy level increases at a higher thermalization degree.
So the net scattering rate of thermalization factor from 0.5 to 1.0 decreases.
Meanwhile, due to the constant oscillation of energy level, Relaxation time has large fluctuations.
Because the materials with high density of states, the DOS of NiSi$_{2}$  and CoSi$_{2}$  within 0.5 eV near the Fermi level is about 0.4-0.6 $eV^{-1}$ (shown in Fig.\ref{NiCoSiDOS} (a) and (b)), which is much higher than that of Si (0-0.1 $eV^{-1}$, Fig.\ref{NiCoSiDOS} (c)).
Therefore, the number of electrons at the Fermi level of NiSi$_{2}$  and CoSi$_{2}$  primitive cell (0.0089 and 0.01068) is much higher than that in the Si conduction band ($2.648 \times 10^{-4}$) under heavy doping (assuming $E_F$ on CBM).
For the same degree of thermalization, the number of carriers in the NiSi$_{2}$  and CoSi$_{2}$  energy windows will be larger than that in the Si.
Fig.\ref{NiCoScattering} (a) and (b) present the increase in number of electrons in the energy windows of NiSi$_{2}$  and CoSi$_{2}$ over time.
It can be found that the number of electrons in the same energy window of NiSi$_{2}$  and CoSi$_{2}$  is 1 to 2 orders of magnitude higher than that in Si (Fig.\ref{Si_Scattering}(a)).
Moreover, the scattering time of thermalization rising edge was extracted (Fig.\ref{NiCoScattering}(c) and (d)).
It was found that the scattering time to thermalization factor of 0.5 was still in the order of sub-100 fs.
By calculating the scattering rate, the curves of the scattering rate changing with the energy window as shown in Fig.\ref{NiCoScattering} (e) and (f).
It can be found that the scattering rate of metallic silicide decreases exponentially with the increase of the energy window at the high energy level, and the scattering rate from partial thermalization to complete thermalization is slightly lower, i.e., similar feature as Si.
However, its scattering rate is higher than that in Si, reaching 3-4 x $10^{33}s^{-1}cm^{-3}$ near the E = 0 eV, which is nearly 2 order of magnitude higher than that in Si (6 - 8 x $10^{31}s^{-1}cm^{-3}$).
Therefore, the scattering rate in the metal may have a more significant effect on the thermalization of the cold source than that of Si.

\subsection{Discussion on Cold carrier transport process with thermalization}
The physical picture of carriers from low energy to high energy calculated by NAMD is caused by setting the initial occupation distribution of the lowest energy level, so it is a dynamic evolution process with time.
However, the actual device will not transit from an initial carrier distribution to the carrier distribution after thermalization at the DC operating point, and the current generated by thermalization and the leakage current of the channel will be in dynamic equilibrium, as shown in Fig.\ref{ThermalTransport}.
Taking the N-type MOSFET as an example, the cold source injects low-energy electrons, and the channel leakage of the cold source device is generated by the scattering of cold carriers to a high energy level in the thermalized region.
When the current generated by thermalized scattering is greater than the current under the original channel thermal balance, the cold source filtering effect fails.
When the current generated by thermalized scattering is less than the current under the channel thermal equilibrium, the leakage current is limited by the injection of the source and the current generated by thermalization.
At this time, the cold source cuts off the high-energy hot electron current.
The cold source injection is not the limiting factor of leakage current because the resistance of the cold source injection junction is very small and meets the requirement of high on-state current.
Because the scattering rate from low energy to high energy decays exponentially with the increase of the energy window, the thermal scattering rate under the channel barrier is much higher than that near the channel barrier. T
he leakage current ultimately depends on the rate of scattering from the energy level near the barrier to above the barrier.
The energy range above the barrier here corresponds to the energy window range calculated by NAMD, and different energy windows represent different barrier heights.
The scattering rate $S_E$ can be qualitatively expained from theory, and the scattering rate for the trasition to the K-state is given by $W(k',k)f(k')[1-f(k)]-W(k',k)f(k[1-f(k')])$ integrating in the $k'$ space.
While $W(k',k)$ is the transition rate from the k' state to the k state, calculated by the Fermi Golden Rule,
\begin{equation}
	W(k',k)=\frac{2\pi}{\hbar}|M_{k',k}|^2 \delta(\hbar\omega_k'-\hbar\omega_k\pm\hbar\omega_q)
\end{equation}
\begin{equation}
	M_{k',k}=<\Psi_k|H'|\Psi_k'>
\end{equation}
where $\omega_k$ is the angular frequency, $M_{k',k}$ is the transition matrix element, H' is the time-dependent perturbation potential, and $\omega_q$is the angular frequency of the perturbation potential.
For the scattering problem under the electron-phonon coupling, the perturbation potential introduced by the lattice vibration determines the transition rate.
$W(k',k)$ reflects the strength of the electro-acoustic coupling. Meanwhile, the scattering rate is also related to the density of States and the occupation probability of the initial and final States.
The density of States is integrated over k' space, and the more initial States satisfy the transition condition, the greater the rate of scattering to K, so the scattering rate and the density of States are positively correlated \cite{fischetti1988monte,ponce2016epw}.
The occupation probability reflects the occupation of the initial state k' and the final state k.
The higher the occupation probability of the initial state, the lower the occupation probability of the final state, the higher the transition probability of the initial state to the final state.
For the reverse process, the lower the transition probability of scattering from k to k' state.
In the energy relaxation process of carrier thermalization, on the one hand, due to the exponential decay of the occupation probability at higher energy, the scattering rate at high energy will decrease exponentially with the occupation probability.
On the other hand, at the beginning of thermalization, the occupation of the initial state is high, and the occupation of the final state approaches zero.
Therefore the scattering at a low degree of thermalization is basically from the initial state at low energy to the final state at high energy, and the rate of scattering from high energy to low energy in the inverse process is very small.
When the high energy level is partially thermalized, the scattering process from low energy to high energy level is weakened while the inverse process scattering is enhanced.
Finally, the scattering between the high and low energy levels is equal, and the thermal equilibrium is reached.
Therefore, the scattering time of the high energy window in NAMD from thermalization at 0 to partial thermalization ($\alpha$= 0.5) is lower than the time from partial thermalization to full thermalization.
In conclusion, the main factors affecting thermalization are the density of states and the electron-phonon coupling intensity related to materials and structures.
The speed of thermalization can be preliminatively measured by scattering time, which represents the speed of thermal equilibrium restoration of adjacent energy states in the system.
The scattering time of thermalization to degree 0.01/0.1 in Si is about 10/40 fs, while the scattering time to degree 0.1 in NiSi$_{2}$  and CoSi$_{2}$  is about 30 fs.
It can be seen that the electron-phonon coupling has no great influence on the scattering process at low thermalization degree.
From Fig.\ref{NiCoThermalization}, the density of state of NiSi$_{2}$  and CoSi$_{2}$  is much higher than that of Si, and it can be qualitatively concluded that thermalization in NiSi$_{2}$  and CoSi$_{2}$ is more serious than that in Si.
Furthermore, based on the energy relaxation scattering time of metallic silicide and silicon obtained by NAMD, the influence of thermalization on the transport characteristics of cold source FET can be calculated in the Boltzmann transport equation framework, which is beyond the scope of this study.

\section{Conclusion}
In this work, DFT based non-adiabatic molecular dynamics (NAMD) is used to study the band structure, density of states and carrier dynamics of bulk Si, NiSi$_{2}$  and CoSi$_{2}$.
Based on our calculation process, the thermalization process of carriers under electron-phonon coupling is analyzed, and the dependence of thermalization factor, relaxation time, scattering time and scattering rate on energy level position have been illustrated.
Under the influence of lattice vibration, cold carrier shifts between energy levels, gradually evolving from low energy occupation to high energy occupation, realizing energy relaxation process, and finally reaching the equilibrium distribution, i.e., exponential decreasing with energy.
The initial thermalization time of high level depends on the thermalization of low level, so the relaxation time increases gradually with the increase of energy level.
The scattering rate in the energy window decreases exponentially with the increase of the energy level, so the thermal leakage of the cold source channel is mainly determined by the thermal scattering rate near the barrier.
Above all, the scattering rate of NiSi$_{2}$  and CoSi$_{2}$  is nearly 2 orders of magnitude higher than that of Si due to the high DOS near the Fermi level, which means the scattering rate of metallic silicide will cause more significant degradation to the thermal leakage of cold source region.
In brief, for silicon-based devices, metallic materials with low mismatch, low density of state and low electron-phonon coupling play a key role to achieve high performance and steep subthreshold device.
	
\section*{Acknowledgements}
This work is supported by the Ministry of Science and Technology (Grant No. 2021YFA1200502), the National Natural Science Foundation of China (Grant No. 12174423, 62174040), and the 13th batch of outstanding young scientific and Technological Talents Project in Guizhou Province [2021]5618.
The authors declare no conflict of interest.


	
\bibliographystyle{apsrev4-1}
\bibliography{IEEEabrv}
	
\end{document}